\documentclass[12pt,epsf]{aastex}
\usepackage{natbib}

\slugcomment{ }
\shorttitle{SN 2014j}
\shortauthors{ et al.}
\usepackage{epstopdf}
\usepackage{graphicx}
\usepackage{amsmath}
\usepackage[caption=false]{subfig}
\usepackage{float}
\newcommand\numberthis{\addtocounter{equation}{1}\tag{\theequation}}

\begin{document}
\title{INTERSTELLAR-MEDIUM MAPPING in M82 THROUGH LIGHT ECHOES 
AROUND SUPERNOVA 2014J} \shorttitle{}
\author{Yi Yang\altaffilmark{1}, 
        Lifan Wang\altaffilmark{1,2},
        Dietrich Baade\altaffilmark{3}, 
        Peter.~J. Brown\altaffilmark{1},
        Misty Cracraft\altaffilmark{4}, 
        Peter A. H\"oflich\altaffilmark{5}, 
        Justyn Maund\altaffilmark{6}, 
        Ferdinando Patat\altaffilmark{3},
        William~B. Sparks\altaffilmark{4}, 
        Jason Spyromilio\altaffilmark{3},
        Heloise F. Stevance\altaffilmark{6}, 
        Xiaofeng Wang\altaffilmark{7}, 
        J. Craig Wheeler\altaffilmark{8}}

\altaffiltext{1}{George P. and Cynthia Woods Mitchell Institute for 
Fundamental Physics $\&$ Astronomy, Texas A. $\&$ M. University, 
Department of Physics and Astronomy, 4242 TAMU, College Station,
TX 77843, USA, email: ngc4594@physics.tamu.edu}
\altaffiltext{2}{Purple Mountain Observatory, Chinese Academy 
of Sciences, Nanjing 210008, China}
\altaffiltext{3}{European Organisation for Astronomical Research 
in the Southern Hemisphere (ESO), Karl-Schwarzschild-Str. 2, 85748 
Garching b.\ M{\"u}nchen, Germany}
\altaffiltext{4}{Space Telescope Science Institute, Baltimore, 
MD 21218, USA}
\altaffiltext{5}{Department of Physics, Florida State University, 
Tallahassee, Florida 32306-4350, USA}
\altaffiltext{6}{Department of Physics and Astronomy, University of 
Sheffield, Hicks Building, Hounsfield Road, Sheffield S3 7RH, UK}
\altaffiltext{7}{Physics Department and Tsinghua Center for 
Astrophysics (THCA), Tsinghua University, Beijing, 100084, China}
\altaffiltext{8}{Department of Astronomy and McDonald Observatory, 
The University of Texas at Austin, Austin, TX 78712, USA}

\begin{abstract}
We present multiple-epoch measurements of the
size and surface brightness of the light echoes from supernova (SN)
2014J in the nearby starburst galaxy M82.
$Hubble\ Space\ Telescope\ (HST)$ ACS/WFC images were taken $\sim277$
and $\sim416$ days after $B$-band maximum in the filters F475W, F606W,
and F775W.  Observations with $HST$ WFC3/UVIS images at epochs
$\sim216$ and $\sim365$ days \citep{Crotts_2015} are included for a more
complete analysis.  The images reveal the temporal evolution of at
least two major light-echo components.  The first one exhibits a
filled ring structure with position-angle-dependent intensity.  This
radially extended, diffuse echo indicates the presence of an
inhomogeneous interstellar dust cloud ranging from $\sim$100 pc to
$\sim$500 pc in the foreground of the SN.  The second echo component
appears as an unresolved luminous quarter-circle arc centered on the 
SN. The wavelength dependence of scattering measured in different dust
components suggests that the dust producing the luminous arc favors
smaller grain sizes, while that causing the diffuse light echo may 
have sizes similar to those of the Milky Way dust.  Smaller grains can 
produce an optical depth consistent with that along the supernova-Earth 
line of sight measured by previous studies around maximum light.  
Therefore, it is possible that the dust slab, from which the luminous 
arc arises, is also responsible for most of the extinction towards 
SN~2014J. {The optical depths determined from the Milky Way-like 
dust in the scattering matters are lower than that produced by the dust 
slab.}
\end{abstract}


\section{Introduction \label{intro}}
Interstellar extinction caused by dust affects most 
astronomical observations. Light traversing a certain distribution 
of interstellar medium (ISM) produces an integrated effect on 
extinction.  Extinction traces the dust grains, but also diminishes 
the starlight and limits our ability to interpret the local and 
distant universe. The study of interstellar dust provides insight 
into the properties of the extinction. Since dust is a strong 
coolant, it also plays a critical role in controlling galaxy 
evolution and star formation. 

Observations of interstellar extinction require a beacon shining
through interstellar material. In the Milky Way, a very large number 
of sightlines are available for this purpose, while in external 
galaxies there are few point source beacons bright enough to study 
the local ISM. Supernovae (SNe) are the best, and often only, choice. 
Light echoes provide additional information because they literally
reflect light-scattering properties and do not reach the observer
along exactly the same path. If SNe are nearby, even resolved light 
echoes may be observable.

The extinction (in magnitudes) at a certain wavelength or bandpass, 
$\lambda$, is often expressed as $A_{\lambda} = R_{\lambda} \times 
E(B-V)$. The `total-to-selective' extinction $ R_V = A_V / E(B-V)$ 
depends on the properties of the dust along the line of sight and 
can be derived by comparing the observed $E(\lambda - V)$ with the 
extinction curves given by \citet{Cardelli_etal_1989}.
{The observed wavelength dependence of }
interstellar extinction contains information on both the size and 
composition of the grains. The value of $R_V = 3.1$ 
\citep{Cardelli_etal_1989} has been often considered the Galactic 
standard, but with a range from 2.2 to 5.8 \citep{Fitzpatrick_1999} 
for different lines of sight. There is increasing evidence that 
extinction curves towards Type Ia SNe exhibit a steeper wavelength 
dependence ($R_V \textless 3$, 
{see \citealp{Cikota_etal_2016} for a summary on $R_V$ results 
of earlier studies}).
\citet{Patat_etal_2007} reported the 
detection of circumstellar material (CSM) in the local environment 
surrounding the Type Ia supernova SN~2006X in the nearby galaxy M100. 
\citet{Wang_2005}, \citet{Patat_etal_2006}, and \citet{Goobar_2008},  
show that the scattered 
light from CSM tends to reduce the value of $R_{\lambda}$ in the 
optical. The effect on $R_V$ and the light curve shape, however, also 
depends on the geometrical configuration and dust-grain properties 
(\citealp{Amanullah_Goobar_2011, Brown_etal_2015}). It is of critical 
importance to understand whether the low $R_V$ values are caused by 
(1) systematic differences from extragalactic environments, or (2) 
inhomogeneities in the vicinity of the SN-Earth direct line of sight 
(DLOS), or (3) modifications by CSM scattering. 

The most reliable approach in determining the extinction is the 
`pair method' --- comparing spectrophotometry of two sources with 
the same spectral energy distribution, one of which has negligible 
foreground extinction. Extragalactic reddening can be measured by 
comparing observed Type Ia SNe to a zero-reddening locus (e.g.,
\citealp{Riess_etal_1996_dust, Phillips_etal_1999}). However,
information acquired through this pair method is limited to single
sightlines. Photons scattered by dust travel a slightly different path
compared to the directly transmitted light. Therefore, scattered
photons provide chances to test the scattering properties of the dust
in a bi-dimensional space.

\subsection{Light echoes}
{Light echoes are from scattered light of a transient event arise from 
dust clouds.} Here we 
consider the case of a SN and CSM/ISM.  Because of the high initial 
brightness of SNe, searches for late-time off-source flux excesses 
have been the main approaches to detect light echoes residing close 
to the SNe, i.e., the slowly fading light curves of SN~1991T 
\citep{Schmidt_etal_1994, Sparks_etal_1999}, SN~1998bu 
\citep{Cappellaro_etal_2001}, and SN~2006X 
\citep{Wang_etal_2008}.  Outside the solar system, spatially resolved 
light echoes have been rare events.  The first one reported arose 
around Nova Persei 1901 \citep{Kapteyn_1901, Ritchey_1901}, followed 
by Nova Sagittarii 1936 \citep{Swope_1940}.  Echoes were also found from 
the Galactic Cepheid RS Puppis \citep{Havlen_1972} and, with {\it HST} 
angular sampling, from the eruptive star V838 Monocerotis 
\citep{Bond_etal_2003}.  \citet{Vogt_etal_2012} reported the detection
of an infrared echo near the Galactic supernova remnant Cassiopeia A. 
Additionally, spectroscopic observations of nearby light echoes provide 
unique opportunities to probe the progenitor properties of historical 
transients \citep{Rest_etal_2008a, Davidson_etal_2012} and in some cases 
the three-dimenisonal structure of the explosion. For instance, an ancient 
eruption from $\eta$ Carinae \citep{Rest_etal_2012}, asymmetry in the 
outburst of SN~1987A \citep{Sinnott_etal_2013} and Cassiopeia A 
\citep{Grefenstette_etal_2014}. In recent years, the 
number of light echoes from extragalactic SNe has grown rapidly, 
mostly thanks to {\it HST}. Table \ref{Table_1} provides an overview 
of the events recorded to date, updated from Table 1 of 
\citet{VanDyk_2015}. 

Photons from spatially-resolved light echoes travel a slightly 
different path compared to the DLOS from the SN to Earth.  Therefore, 
observations of a resolved light echo around a nearby SN provide a 
unique opportunity to measure the extinction properties of the dust 
along the DLOS and the scattering properties of the echo-producing 
dust independently and simultaneously.  As the SN fades, outer echoes 
(echoes with larger angular diameter) associated with ISM at large 
distances to the SN will become less contaminated by its 
bright light, and any inner echoes associated with ISM at small 
distances to the SN, and even the CSM, will become 
detectable.  The expansion with time of the light echoes maps
out the 3D structure of ISM along and close to the line of sight.

\begin{figure}
\epsscale{0.8}
\plotone{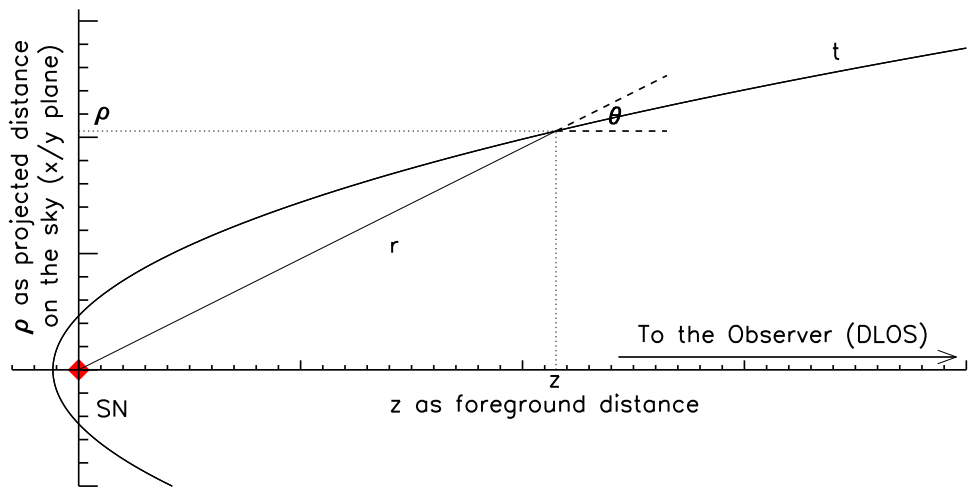}
\caption{Schematic diagram identifying the
  geometrical parameters used in this paper. The paraboloid represents
  the iso-delay light surface at some arbitrary epoch after the
  supernova explosion.  The observer located along the $z$-axis and
  beyond the right edge of the diagram would see light echoes in the
  $x$-$y$ plane (the $y$ is perpendicular to the drawing).  The SN
  is located at the origin and $\theta$ denotes the scattering angle.
\label{fig_1}} 
\end{figure}

Detailed introductions to the relation between 2-dimensional light 
echoes and 3-dimensional scattering dust distributions has been given 
in various studies \citep{Chevalier_1986,Sparks_1994,Sugerman_2003,
Tylenda_2004, Patat_2005}.  Here, we just briefly define the geometry 
used through this paper, also shown in Figure \ref{fig_1}, which 
considers the SN event as an instantaneous flash of radiation. 
The locus of constant light travel time is an ellipsoid with the 
supernova at one focus which we refer to as an iso-delay surface. The 
ellipsoid grows with time as the light propagates in space.

The angular radius of the light echo ($\alpha$) can be easily measured
in two-dimensional images.  
{The SN is centered at the origin of the plane, the  $x$ and $y$ give the coordinates 
of the scattering materials in the plane of the sky}. The projected distance 
($\rho = \sqrt{x^2 + y^2}$) of scattering material to the SN perpendicular 
to the DLOS is related to the distance ($D$) to the SN as 
$\mathrm{\tan}\alpha=\rho/D$, $\phi$ gives the position angle (PA). 
Because $D$ is significantly larger 
compared to other geometric dimensions, the light echo can be very well
approximated by a paraboloid, with the SN lying at its focus. $\rho$ can 
be obtained by 
\begin{equation}
\rho=\sqrt{ct(2z+ct)}.
\label{eqn_1}
\end{equation} 
where $t$ is the time since the radiation burst, $z$ gives the foreground
distance of the scattering material along the line of sight, and $c$
denotes the speed of light.  The distance $r$ of the scattering material
from the SN is:
\begin{equation}
r=\frac{1}{2} \bigg{(} \frac{\rho^2}{ct} + ct \bigg{)}
\label{eqn_2}
\end{equation}
The scattering angle can be obtained from: 
$\mathrm{cos}\ \theta(\rho,t)=z/(z+ct)$, or, $\mathrm{tan}\theta=\rho /z$. 

\subsection{Supernova 2014J in M82}
The nearby Type Ia SN~2014J in M82 
{(3.53$\pm$0.04 Mpc, \citealp{Dalcanton_etal_2009})}
offers the rare opportunity to 
study the physical properties and spatial distribution of dust
particles along {\it and close to} the {DLOS and as well in the
vicinity of the SN}. 
SN~2014J suffers from heavy extinction 
{($A_V$ = 2.07$\pm$0.18, \citealp{Foley_etal_2014})} 
and is located behind a 
large amount of interstellar dust \citep{Amanullah_etal_2014}. 
Additionally, the absorption profiles of Na and K lines from 
high-resolution spectroscopy exhibit more than ten extragalactic 
absorption components, indicating the extinction along the DLOS 
is caused by the combined presence of a large number of distinct 
interstellar dust clouds along the DLOS \citep{Patat_etal_2015}. 
SN~2014J was discovered on Jan 21.805 UT by \citet{Fossey_etal_2014}. 
Later observations constrained the first light of the SN to 
Jan.\ 14.75 UT \citep{Zheng_etal_2014, Goobar_etal_2014}.  

SN~2014J reached its $B-$band maximum on Feb. 2.0 UT (JD 2,456,690.5)
at a mgnitude of 11.85$\pm$0.02 \citep{Foley_etal_2014}.  Continuous 
photometric and spectroscopic observations through late phases have
been made by various groups \citep{Johansson_etal_2014,
  Lundqvist_etal_2015, Bonanos_etal_2016, Srivastav_etal_2016, 
  Porter_etal_2016, Sand_etal_2016}.

There is clear evidence that the strong extinction measured from 
SN~2014J is caused primarily by interstellar dust 
\citep{Patat_etal_2015, Brown_etal_2015}, although a mix of interstellar
and circumstellar dust is also possible \citep{Foley_etal_2014, 
Bulla_etal_2016}. Several independent studies, including photometric 
color fitting from Swift/UVOT and {\it HST} \citep{Amanullah_etal_2014}, 
near-UV/optical grism spectroscopy from Swift UVOT 
\citep{Brown_etal_2015}, {\it HST} STIS spectroscopy and WFC3 
photometry \citep{Foley_etal_2014}, reddening curve fitting near 
the SN maximum using the silicate-graphite model \citep{Gao_etal_2015}, 
as well as optical spectroscopy from \citet{Goobar_etal_2014} found an 
$R_V \sim 1.4$ towards SN~2014J. Moreover, ground-based broad-band 
imaging polarimetry \citep{Kawabata_etal_2014, Srivastav_etal_2016} 
and spectropolarimetry \citep{Patat_etal_2015, Porter_etal_2016} have 
shown that the polarization peak due to interstellar dust extinction 
is shortward of $\sim0.4\mu m$, which indicates that this line of sight 
has peculiar Serkowski parameters (see \citealp{Patat_etal_2015}). This 
polarization wavelength dependence can be interpreted in terms of a 
significantly enhanced abundance of small grains \citep{Patat_etal_2015}. 
Models considering both interstellar 
dust and circumstellar dust simultaneously and fitted to observed 
extinction and polarization \citep{Hoang_2015} find that a significant 
enhancement (w.r.t.\ the Milky Way) in the total mass of small grains 
($\textless$ 0.1 $\mu m$) is required to reproduce low values of $R_V$. 
Multiple time-invariant \ion{Na}{1} D and \ion{Ca}{2} H\&K absorption 
features as well as several diffuse interstellar bands (DIBs) have also 
been identified \citep{Graham_etal_2015, Jack_etal_2015}. Those are 
most likely associated with multiple dust components of interstellar 
material along the DLOS.  

The nature (amount and distribution) of circumstellar material is of 
interest when probing the possible diversity of progenitors of type Ia 
SNe and for accurately correcting the extinction when using type Ia SNe 
as standard candles. 
\citet{Johansson_etal_2014} find no evidence for heated dust
in the CSM of SN~2014J with r$\textless 10^{17}$ cm ($\sim39$ light 
days). \citet{Graham_etal_2015} reported variable interstellar \ion{K}{1} 
lines in high-resolution spectra, which may form about 10 light years
($\sim10^{19}$ cm) in front of the SN. 

{The extremely dusty environment in M82 and its relative proximity} to
Earth lead to the expectation of complex and evolving light echoes if
SN~2014J exploded inside the galactic disk.  In fact, \Citet{Crotts_2015}
discovered the first light echoes surrounding SN~2014J in {\it HST}
images from September 5 2014, 215.8 days past $B$-band maximum
light (referred to as +216 d hereafter) on JD = 2456690.5
\citep{Foley_etal_2014}. 
The echo signal tends to be associated with pre-explosion nebular 
structures in M82 \citep{Crotts_2015}. 

In the following, we present the evolution of multiple light echoes of
SN~2014J as revealed by new {\it HST} ACS/WFC multi-band and multi-epoch
imaging around $\sim$277 days and $\sim$416 days past $B$-band maximum 
(referred to as +277 d and +416 d below). We will also qualitatively 
discuss similar archival WFC3/UVIS images obtained on +216 d and +365 d. 

\section{Observations and Data Reduction}
Late-time observations of the light echoes around SN~2014J discussed
in this paper result from a {\it Hubble Space Telescope (HST)} Wide
Field Camera 3 UVIS channel ({\it HST} WFC3/UVIS) program ($\#$13626; 
PI:Crotts) to observe properties of the light echoes and progenitor
environment around SN~2014J and an Advanced Camera for Surveys/Wide
Field Channel ({\it HST} ACS/WFC) program ($\#$13717; PI: Wang) to 
probe the dusty environment surrounding SN~2014J in M82.  A log of
observations is assembled in Table~\ref{Table_2}.

\begin{figure}[!htbp]
\epsscale{1}
\plotone{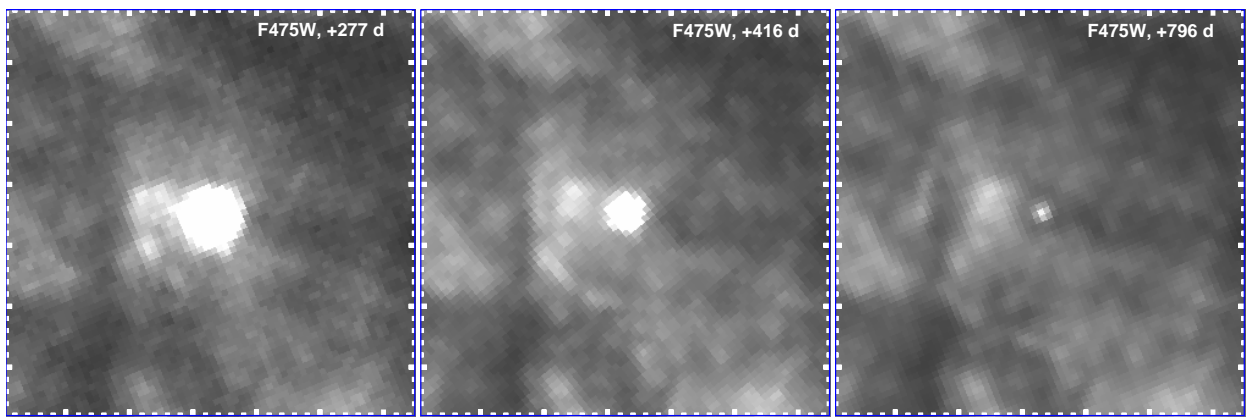}
\caption{{\it HST} ACS/WFC F475W images of SN~2014J obtained at
  different epochs as labeled.  Each square measures 3.2$\arcsec$= 54 pc
  along its sides (North is up, East is left).  The distance between little
  tick marks corresponds to 0.1$\arcsec$.  
\label{fig_2}} 
\end{figure}

We use bright HII regions to align exposures in different filter
combinations and epochs through {\it Tweakreg} in the {\it
  Astrodrizzle} package \citep{Gonzaga_etal_2012}.  Observations
obtained with three polarizers are needed to calculate the Stokes
vectors, but the intensity maps (Stokes I) are the only input to this
analysis.
\begin{align*}
I = \frac{2}{3} [r(POL0) + r(POL60) + r(POL120)],  
\numberthis \label{eqn_3}
\end{align*}
where $r(POL0)$, etc.\ are the count rates in the images obtained
through the three polarizers. 
Figure \ref{fig_2} shows the field around SN~2014J. 

{We perform background subtraction} to better reveal the faint and 
time-variant light echo signals.  For observations on +277 d 
and +416 d with {\it HST} ACS/WFC and filters F475W, F606W, and F775W, 
we found no pre-SN Hubble images of the region through filters 
consistent with our observations.  The most recent {\it HST} images of 
SN~2014J obtained on April 8 2016, (+796 d) with the same photometric 
and polarimetric filter combinations were subtracted from the 
observations on +277 d and +416 d.  For the observations on +216 d and 
+365 d with {\it HST} WFC3/UVIS in passbands F438W, F555W, and F814W, 
pre-SN images obtained on March 29 2006 (program $\#$10776; 
PI:Mountain) with {\it HST} ACS/WFC in the F435W, F555W, and F814W were 
used as background templates, respectively.  For each band, the 
background templates were scaled and subtracted from the intensity map.

\begin{figure}[!htbp]
\epsscale{1}
\plotone{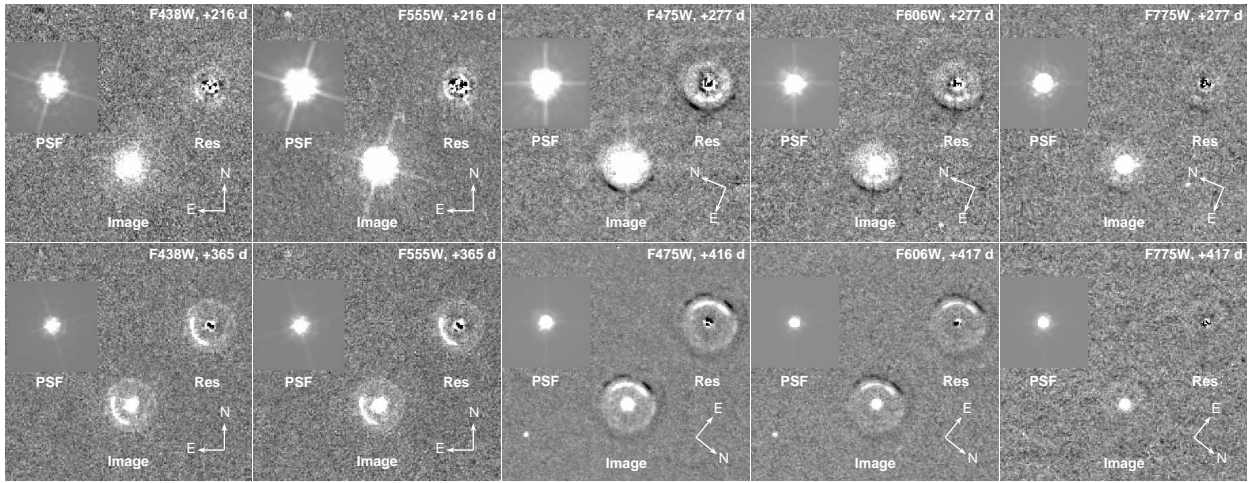}
\caption{Background-subtracted images of the SN (`Image'), the TinyTim
  PSF \citep{Krist_1993, Krist_etal_2008}, and the residuals around
  the SN after PSF subtraction (`Res').  Background structures in
  F438W and F555W were removed by subtracting scaled pre-SN archival
  F435W and F555W HST images.  Background in F475W, F606W, and F775W
  was corrected for by subtracting the respective most recent +796 d
  image; therefore, the +796 d echoes appear as negative structures.
  Note the different orientations.
\label{fig_3}} 
\end{figure}

The resulting images (Figure \ref{fig_3}) clearly reveal the shape of 
the light echoes around SN~2014J. Negative 
signals (black in Figure \ref{fig_3}) represent the light echoes 
on +796 d while positive (white) signals trace the light echoes on 
+277d and +416 d, respectively.  In each subpanel of Figure \ref{fig_3},
we show the light echoes with background removed (labeled `Image' at
bottom), the scaled and distortion-corrected PSF (labeled `PSF' on the
left), and the residual around the SN after PSF subtraction (labeled
`Res' on the right). Point-spread functions (PSF) appropriate to the SN 
position were generated for each bandpass and epoch with TinyTim 
\citep{Krist_1993, Krist_etal_2008}. 
The upper row displays the observations at
earlier epoches (+216 d for F438W and F555W, +277 d for F475W, F606W,
and F775W), and the lower row depicts the observations at later epochs
(+365 d for F438W and F555W, +416 d for F475W, F606W, and F775W).  For
better visibility, Figure \ref{fig_4} provides a zoom-in of the
PSF-subtracted images (`Res') in each panel of Figure \ref{fig_3}.

\section{Analysis and results}
\subsection{Total flux of the SN}
Photometry of SN~2014J at four epochs was performed in the background
subtracted images described above, {and shown in Table \ref{Table_3}}.
Measurements were made with a circular aperture of 0.4$\arcsec$ (8
pixels in the ACS/WFC FOV and 10 pixels in the WFC3/UVIS FOV) in the 
WFC3/UVIS F438W, F555W, F814W images from +216 d, and the F438W and 
F555W images from +365 d. We applied aperture corrections according to 
\citet{Hartig_2009} and \citet{Sirianni_etal_2005} to estimate the 
total flux from SN~2014J.  The photometric uncertainties in Table 
\ref{Table_3} include the Poisson noise of the signal, the photon noise 
of the background, the readout noise contribution (3.75 electrons/pixel 
for ACS/WFC), and the uncertainties in aperture corrections. These 
quantities were added in quadrature. 
The magnitudes are presented in the Vega system with zero points from 
the CALSPEC
\footnote[1]{http://www.stsci.edu/hst/observatory/cdbs/calspec.html,
  or http://www.stsci.edu/hst/acs/analysis/zeropoints/\#vega and
  http://www.stsci.edu/hst/wfc3/phot\_zp\_lbn} archive.  
The total flux of the source within the aperture equals the product
Total Counts $\times$ PHOTFLAM, where PHOTFLAM is the inverse
sensitivity (in erg cm$^{-2}$ sec$^{-1}$ $\rm{\AA^{-1}}$ and representing a
signal of 1 electron per second).  For WFC3/UVIS images, we adopted
the values of the PHOTFLAM keyword in the image headers.  However, for
the ACS/WFC polarizer images, which were corrected for the throughputs
of the polarizers to generate the intensity maps, we discarded the
default PHOTFLAM values.  Instead, we adopted the most up-to-date
PHOTFLAM values in the ACS filter bands for images obtained without
polarizers \citep{Bohlin_2012}.  This is required by the mismatch
between (i) the polarizer throughput curves used by
SYNPHOT\footnote[2]{\footnotesize http://www.stsci.edu/institute
  /software\_hardware/stsdas/synphot} for unpolarized sources and (ii)
the values found by comparing unpolarized sources in both the
polarizing and non-polarizing filters \citep{Cracraft_Sparks_2007}.
Therefore, the PHOTFLAM keywords in ACS/WFC polarized images are not
applicable to intensity maps derived from polarized images.
Polarization properties of SN~2014J will be discussed in a separate
paper (Yang et al., in prep.).

\subsection{Residual images}
\begin{figure}[!htbp]
\epsscale{1}
\plotone{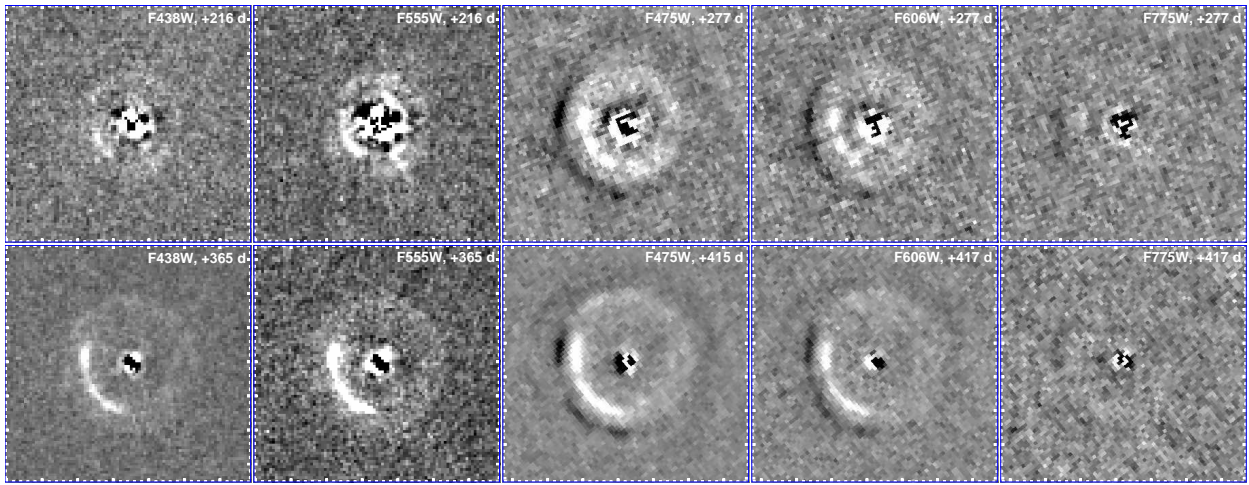}
\caption{A zoom-in view of the background-corrected light echoes shown
  in Fig.\ \ref{fig_3}.  North is up and East is left.  The distance
  between each little tickmark is $0.1 \arcsec$. 
  Each square measures 3.4$\arcsec$= 58 pc along its sides. 
  The diffuse and radially extended light
  echo profiles can be clearly identified in all panels except for
  F438W (+216 d) and F775W (all epochs).  Note the uneven signal
  distribution with position angle in the rings and the consistency of
  the overall patterns at different epochs.  A luminous arc is visible
  in the lower left quadrant and not resolved in the radial direction.
  This is at variance with the appearance of the complete, radially 
  diffuse rings.  
\label{fig_4}} 
\end{figure}

Two main echo components are evident. 
In Figure \ref{fig_4} we show a luminous quarter-circle 
arc and a diffuse ring at angular distance larger than 
0.3$\arcsec$ from the SN. Closer to the SN, uncertainties in the PSF 
correction prevent reliable detections.  On +277 d, the most notable 
features of the light echoes in 
F475W are three luminous clumps at angular radius $\alpha=0.60\arcsec$ 
and PAs 80$^{\circ}$, 120$^{\circ}$, and 150$^{\circ}$, measured 
from north (0$^\circ$) through east (90$^\circ$).  These clumpy 
structures are already present on +216 d at the same PAs but appear 
smoother and more extended.  They eventually evolve into a fairly 
continuous luminous quarter-circle arc seen on both +365 d and +416 d 
extending from PA = 60$^{\circ}$-170$^{\circ}$. Images obtained on 
+216 d with F438W and F555W show the luminous arc at angular radii 
$\alpha=0.54\arcsec$ and $\alpha=0.69\arcsec$, 
over roughly the same range in PA, in agreement with 
\citet{Crotts_2015}. However, for the arc we find a foreground distance 
of the scattering material, which ranges from 226 to 235 pc in the 
four epochs (Table \ref{Table_4}) and has a mean value of 228$\pm$7 pc. 
This is different from the foreground distance of $\sim$ 330 pc 
discussed for this prominent echo component by \citet{Crotts_2015}. 
This discrepancy may be due to the difficulties and uncertainties in 
subtracting the PSF in earlier epoch when the SN is still bright, or 
in distinguishing the multiple light echo components identified in 
our multi-epoch data. 

To enable a more quantitative description of the light echoes and
their evolution, we performed photometry on them in backgound-subtracted
images (Figure \ref{fig_4}).  We measured the surface 
brightness of the light echo profile at different radii and over 
different ranges in PA.  Fan-shaped apertures centered on the SN
were used to sample the intensity.  The width in PA of each aperture
is $45^\circ$.  Contrary to the luminous arc, the diffuse echo can 
be seen over the full range in PA from 0$^{\circ}$ to 360$^{\circ}$. 
But it does not exhibit a common radial profile (Figures \ref{fig_5} 
and \ref{fig_6}).

In the following subsections, we will use these measurements to
investigate the evolving profile of the light echoes, conduct
geometric and photometric analyses, and estimate the dust distribution 
and scattering properties responsible for the observed light echoes 
along and close to the DLOS.  A function characterizing the properties 
of the scattering material is constructed to represent the brightness 
evolution of the observed light echoes on +277 d and +416 d. 

\subsection{Geometric properties of the light echoes}
A comprehensive discussion of the formation of light echo arcs is 
available from \citet{Tylenda_2004}.  In the context of this paper, 
it is sufficient to recall that 
{a circular light echo is created from the intersection of the dust 
slab with the iso-delay parabaloid}. 
Any uneven distribution of
material in the slab results in an uneven flux distribution along the
circle, and the light echo may be composed of incomplete arcs.  A dust
slab always produces a (complete or incomplete) circular light echo,
irrespective of its inclination with respect to the line of sight. 
When a dust slab is not perpendicular to the line of sight, 
the center of the light echo circle will not coincide with the SN 
position, and it moves with time.

The luminous arc echo is unresolved with a 
full width at half maximum (FWHM) of the radial profile approximately
that of the SN measured in the same images, 
i.e. $\sim$ 0.1$\arcsec$ (2 pixels). Therefore, we consider the luminous 
arc was formed by a thin dust slab intersecting the line of sight. We 
have fitted circles to the positions of the luminous arc at all available 
epochs. {None of them are significantly decentered from the SN}.  This 
implies that the dust slab producing the arc echo is fairly
perpendicular to the line of sight. {Table \ref{Table_4} summarizes
the geometric properties} measured from the luminous arc.

In addition to the luminous arc, 
a radially extended and diffuse structure is identified, 
which on +277 d is present in F475W and F606W and spread
over $\alpha = 0.40\arcsec$ to $\alpha = 0.90\arcsec$.  This structure 
can also be noticed on +365 d in F438W and F555W (from $\alpha = 0.47
\arcsec$ to $\alpha = 1.03\arcsec$).  It appears more clearly on 
+416/417 d in F475W and F606W (from $\alpha = 0.50\arcsec$ to 
$\alpha = 1.08\arcsec$) because for these observations longer exposure 
times were used.  The epochs of observation and the exclusion of the 
inner 0.3$\arcsec$ limit the foreground distances explored from 
$z=100$ pc to $z=500$ pc.  On +216 d, the diffuse 
component cannot be identified in F438W but is marginally seen in F555W. 
However, the inner and outer radii of the diffuse structure cannot be 
well determined because of uncertainties in the PSF subtraction. The 
diffuse light echo observed on +277 d can be produced by a dust cloud 
intersecting the iso-delay surface over a wide range in foreground 
distance.  The line-of-sight extent of this diffuse dust cloud is 
indicated by the filled profile of the echoes. Therefore, a continuous 
dust distribution over a certain range of foreground distances along 
the line of sight is required. 

\begin{figure}[!htbp]
\epsscale{1}
\plotone{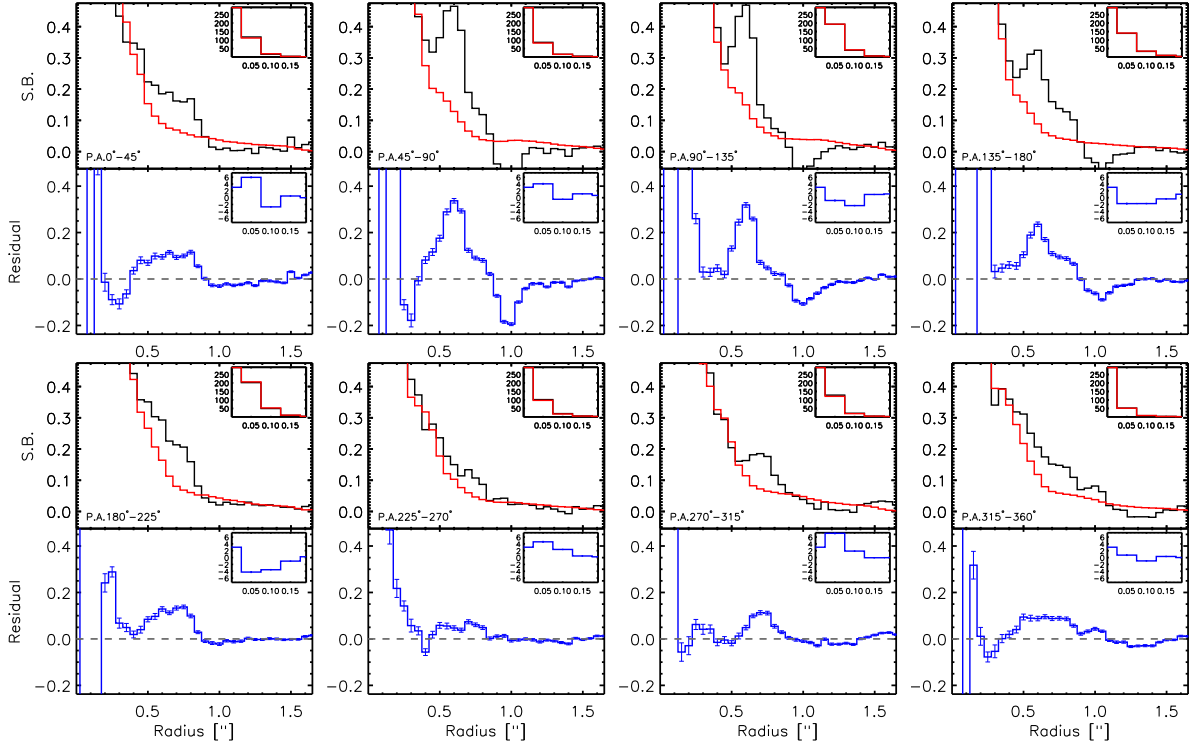}
\caption{F475W band radial surface-brightness profile centered on
  SN~2014J at 277 days after B-band maximum.  Different curves in each
  panel show the surface brightness of the background-subtracted image
  (black), the TinyTim PSF (red), and the residual after PSF subtraction 
  (blue). Each panel depicts a different 45$^\circ$ sector in PA as 
  labeled. The lower subpanels of each pair display the residuals after 
  PSF and background subtraction; the luminous arc at $\sim0.6$ arcsec 
  is prominent in the PA bins from 45$^\circ$ to 180$^\circ$. The 
  diffuse light echoes can be identified at other PAs, by continuous 
  positive signals from the early epoch of +277 d and continuous 
  negative signals due to the subtraction of the intensity map on +796 d. 
  Surface brightnesses are in units of
  $\mathrm{10^{-16}erg\ s^{-1} \AA ^ {-1} arcsec^{-2}}$.  The inserts
  display the radial run of the functions (identified by their colors)
  over the innermost 0.2 arcsec. 
\label{fig_5}} 
\end{figure}

\begin{figure}[!htbp]
\epsscale{1}
\plotone{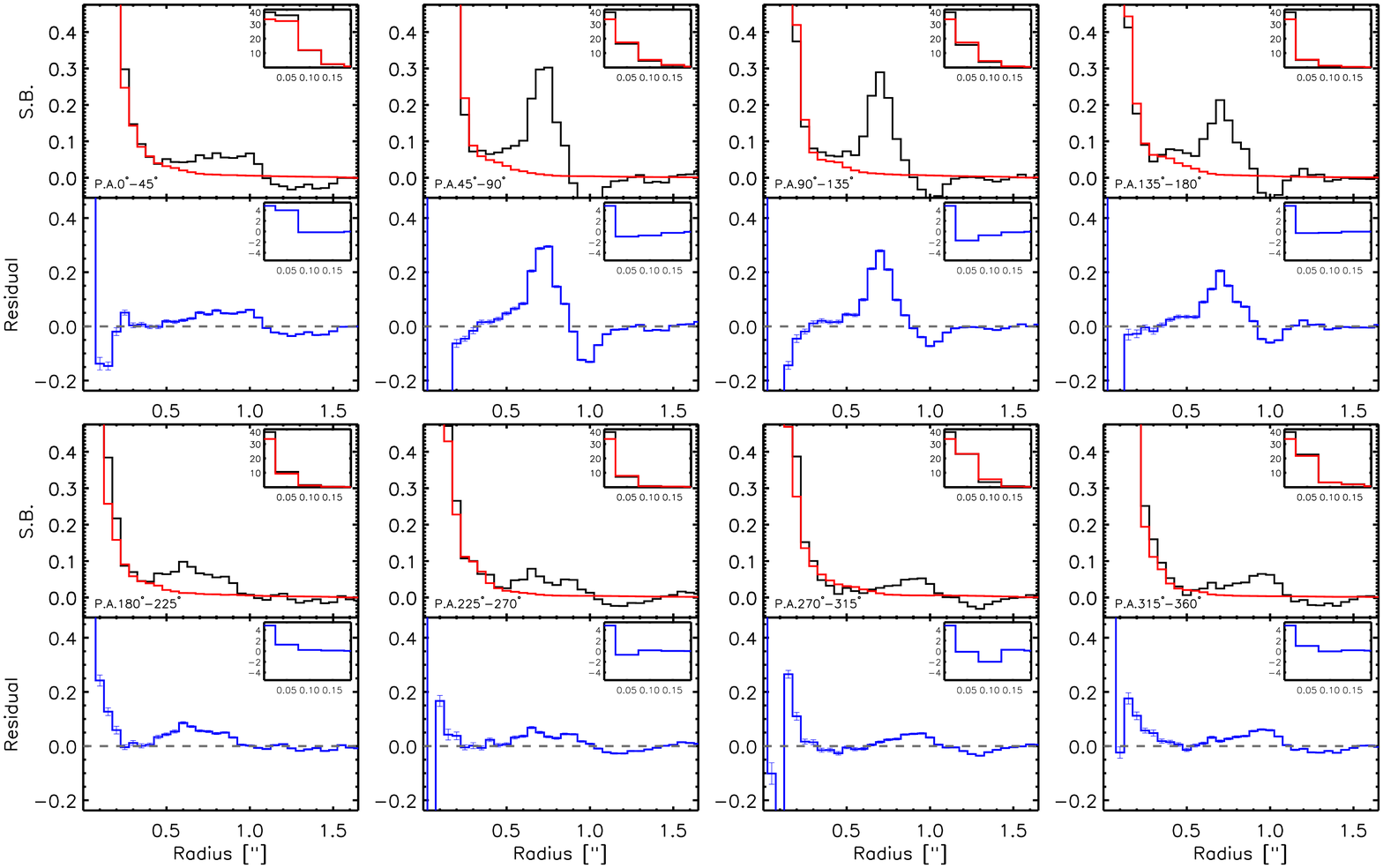}
\caption{Same as Figure \ref{fig_5} except for epoch being +416 d.
\label{fig_6}} 
\end{figure}

In each panel of the radial profiles in Figures \ref{fig_5} and
\ref{fig_6}, the radially-resolved positive flux excesses (on +277 d
and +416 d), and also the radially-extended negative flux due to the
subtraction of the light echo on +796 d, suggest the presence of an
extended and inhomogeneous foreground dust distribution. {Outside the 
$\sim 0.3\arcsec$ region, as discussed earlier}, the imperfect 
PSF subtraction makes the detection of echoes unreliable. The most 
prominent structure with an intensity peak at the second and third
curve near the top in Figure \ref{fig_5} can be seen clearly on +277 d
with an angular radius of $\sim0.60\arcsec$, which at the distance of 
M82 (3.53$\pm$0.04 Mpc, \citealp{Dalcanton_etal_2009}) is at a radius 
$\rho = 10.3$ pc from the SN in the plane of the sky.  
By +416 d, the radius has increased to 
$\sim0.735\arcsec$ or $\rho = 12.6$ pc from the SN.  The scattering 
angles are $2.6^{\circ}$ and $3.2^{\circ}$, respectively.  

\subsection{Light echo mapping of the foreground dust distribution} 
{To our knowledge, and with the exception of SN~1987A in the 
LMC (\citealp{Crotts_1988}, \citealp{Suntzeff_etal_1988})}, 
this is the first radially-extended light echo detected from any SN. 
For epochs discussed in this paper, the diffuse echo component around 
SN~2014J reveals the SN-backlit ISM over $\sim$ 40 pc $\times$ 40 pc 
around the DLOS.  Standard methods for estimating the optical properties 
of the ISM towards the supernova only consider the extinction along the
DLOS. They include the spectrophotometric comparison between the 
observed SN and an unreddened SN or template, and comparing the 
integrated echo flux with the surface brightness calculated from the
scattering properties of various dust models. But the resolved dust 
echoes of SN~2014J and their temporal evolution in the gas-rich and 
very nearby galaxy M82 provide an unprecedented opportunity to do 
better.  In the following, we take advantage of 
this to measure the scattering properties of the ISM at 
different foreground distances and PAs relative to SN~2014J.

We assume that dust scattering follows the Henyey--Greenstein phase 
function \citep{Henyey_Greenstein_1941}: 
\begin{equation}
\Phi(\theta) =   
\frac{1-g^2}{(1+g^2-2gcos\theta)^{3/2}}
\label{eqn_phasefunction}
\end{equation}
where $g=\overline{cos\theta}$ is a measure of the degree of forward
scattering.  With $L_{\lambda} (t)$ as the number of photons emitted per unit
time by the SN at a given wavelength, $F_{\lambda}(t) = L_{\lambda} (t)/4 \pi D^2$ is the
number of photons observed at time $t$.  $D$ is the distance to the SN. 
For the modeling of our observations, t is the time of the light-echo 
observation, $t_e$ denotes the time when photons emitted 
by the SN would be directly observed along the DLOS, and $F_{\lambda} (t-t_e)$ 
is the brightness of the SN at $(t-t_e)$.  At $t$, the photons emitted 
at the same time as $t_e$, but experiencing scattering leading to a 
light echo, arrive at the observer with a time delay $(t-t_e)$.

{For a single short flash of light} of duration $\Delta t_e$
emitted by the SN at $t_e$, $F_{\nu}(t-t_e)=0$ for $t\ne t_e$ and
$\int_{0} ^{t} F_{\nu}(t-t_e) dt_e = F_{\nu} (t-t_e) | _{t=t_e} \Delta
t_e$.  Then, the surface brightness, $\Sigma$, of a scattered-light 
echo at frequency $\nu$ and arising from an infinitely short ($\delta$ 
function) light pulse is given by: 
\begin{equation}
\Sigma_{\nu}^{\delta} (\rho,\phi,t) = n_d Q_s \sigma_d \ 
\frac{F_{\nu}(t-t_e)| _{t=t_e} \Delta t_e}{4 \pi r^2} \ 
\bigg{|}\frac{dz}{dt}\bigg{|} \Phi(\theta)
= n_d Q_s \sigma_d \frac{\int_{0} ^{t} F_{\nu}(t-t_e) dt_e}
{4 \pi r^2} \bigg{|}\frac{dz}{dt}\bigg{|} \Phi(\theta)
\end{equation}
{Where $n_d$ is the volume number density} of the scattering material 
in units of $cm^{-3}$; $Q_s$ is a dimensionless number describing the 
scattering efficiency of the dust grains; 
$\sigma_d$ is the geometric cross-section of a dust grain, 
$\Phi(\theta)$ is the unitless scattering phase function. 
This means that the surface brightness at a certain instance of the 
light echo at $t = t_e + (t-t_e)$ is determined by 
the flux emitted from the SN at $t_e$, together with the local 
geometric properties of the iso-delay surface at $t-t_e$. 

In reality, the SN emission has a finite duration.  $F_{\nu}(t-t_e)$
is no longer a $\delta$ function, and the surface brightness of the
light echo unit at a certain frequency $\Sigma_{\nu}$ is the time 
integral of $F_{\nu}(t-t_e)$ from $0$ to $t$:
\begin{equation}
\Sigma_{\nu} (\rho,\phi,t) = \frac{Q_s \sigma_d}{4 \pi} 
\int_0 ^t \frac{n_d F_{\nu} (t-t_e) d t_e}{r^2} \ 
\bigg{|}\frac{dz}{dt}\bigg{|} \Phi(\theta)
\end{equation}
Recalling that 
\begin{equation}
z = \frac{\rho^2}{2ct} - \frac{ct}{2}
\label{eqn_z}
\end{equation}
one can easily find:
\begin{equation}
\frac{dz}{dt} = -\frac{c}{2} \bigg{(} \frac{\rho^2}{c^2 t^2} + 1 \bigg{)}, \ 
r=z+ct=\frac{ct}{2} \bigg{(} \frac{\rho^2}{c^2 t^2} + 1 \bigg{)}
\label{eqn_r}
\end{equation}
Therefore, 
\begin{equation}
\Sigma_{\nu}(\rho,\phi,t) = \frac{Q_s \sigma_d c}{2 \pi} \int_0 ^t 
\frac{n_d(\rho,\phi,t)}{c^2 t^2 + \rho^2} \Phi(\theta) \ F_{\nu} (t-t_e) dt_e
\label{eqn_lumi}
\end{equation}
Because of the relative proximity of M82, some light echoes around 
SN~2014J are resolved by {\it HST} at late
phases, and each pixel represents the surface brightness of the light
echo multiplied by the physical area covered by the pixel in the sky. 

Therefore, in order to compare the model flux distribution with the
flux in a 2-dimensional image, one needs to integrate the model flux
over the physical depth covered by the pixel.  Since each pixel has 
size $\Delta x \Delta y$, and $\Delta x = \Delta y$, this implies:
\begin{equation}
Im_{\nu}(x,y,t) = \int_{x-\frac{\Delta x}{2}}^
{x+\frac{\Delta x}{2}} \int_{y-\frac{\Delta y}{2}}^
{y+\frac{\Delta y}{2}} \Sigma(x,y,t) dx dy 
\end{equation}

The geometric factor is determined by the radial distance to the SN,
$\rho = \sqrt{x^2 + y^2}$.  Therefore, in the tangential direction inside 
each pixel, we approximate the integration by assuming that $n_d (x,y,t)$ 
is invariant over the angle $\Delta \phi$ subtended by a single pixel. 
Furthermore, the angular size of each ACS/WFC pixel is $0.05\arcsec$. 
At the distance of $D=3.53\pm0.04$ Mpc, the corresponding physical 
pixel size in the sky is:
\begin{equation}
pixscale = (3.53\pm0.04) \ \mathrm{Mpc} \times tan(0.05\arcsec) 
= (0.86\pm0.01) \  pc = \Delta x = \Delta y
\end{equation}

\begin{figure}[!htbp]
\epsscale{0.8}
\plotone{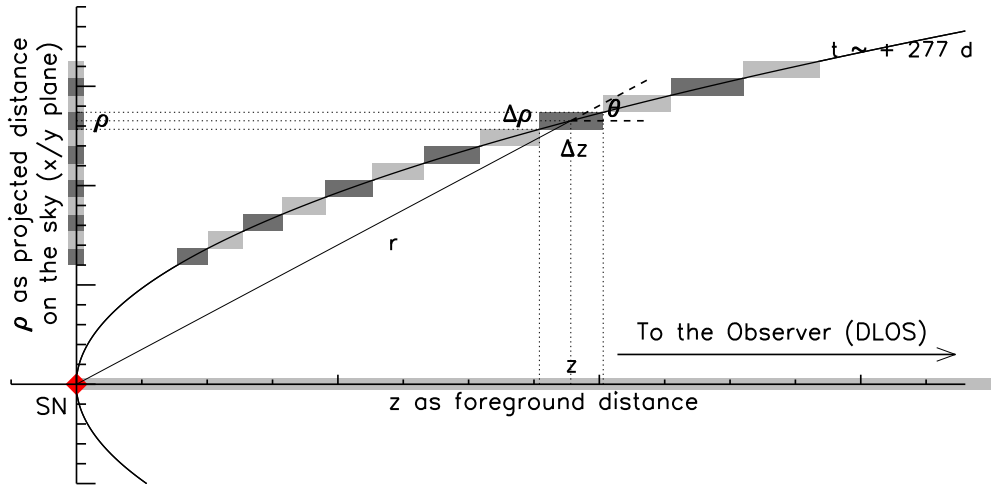}
\caption{
  Schematic diagram from Figure \ref{fig_1} adapted to  
  real scale. The paraboloid represents the iso-delay 
  light surface at $\sim$ 277 d. The gray-shaded squares on 
  the vertical axis indicate the pixelation of echo signals measured from 
  images of this epoch. {Rectangles at the same observed} angular distance
  delineate the range in $z$, over which dust can produce an unresolved
  light echo. Different gray levels only serve to distinguish immediately 
  neighboring pixels. 
\label{fig_7}} 
\end{figure}

Recall the geometric configuration of the iso-delay light surface at 
+277 d presented by Figure \ref{fig_1}.  In Figure \ref{fig_7}, we 
modify this schematic diagram to demonstrate how we use a 2-dimensional 
image to map the ISM in 3D. The gray-shaded fields on the vertical axis 
show the pixelation of the sky view by the camera, with each pixel 
measuring 0.86 $pc$ on both sides. $\Delta z$ is the position-dependent 
line-of-sight extent of the foreground column covered by each pixel. 
Gray-shaded rectangles superimposed to the iso-delay light surface 
mark columns of ISM which would be responsible for 
respective light echoes as projected onto the sky. The fixed size of 
the sky pixels leads to varied lengths of the foreground columns of 
ISM. {If the ISM is homogeneously distributed in the $x/y$ plane}, the 
total per-sky-pixel extinction of the scattering materials as revealed by 
the light echo can be estimated by summing up the extinction along each 
rectangular column of ISM intersecting the iso-delay light paraboloid. 
Comparison of the extinction by the scattering materials to the extinction 
along the DLOS (marked by the gray line on the $z$-axis in Figure \ref{fig_7})
may reveal if they are caused by the same dust mixture and perhaps even the
same dust cloud.  

Now we can compare the intensity map obtained from the observations
with the light echo modeled at each physical position for a given 
time $t$ of the observation as follows:
\begin{equation}
Im_{\nu}(x,y,t) = \frac{\omega C_{ext} c}{2 \pi} \Delta x 
\int_{x-\frac{\Delta x}{2}}^{x+\frac{\Delta x}{2}} dx 
\int_{0}^{t} \frac{n_d (x,y,t)}{c^2 t^2 + \rho^2} 
\Phi(\theta) F_{\nu} (t-t_e) dt_e
\label{eqn_map}
\end{equation}

\subsection{Extinction of the scattering materials}
The optical properties of the dust grains responsible for the light 
echoes around SN~2014J can be deduced within each observed pixel. 
We estimate the extinction properties of the scattering materials 
based on a single-scattering-plus-attenuation approach (see Section 5 
of \citealp{Patat_2005} for more details). 
{Conversions from the intensity map to the number-density 
map ('nd') are presented by Figure 8 based on Equation 12.} 
We follow the sampling in Figures \ref{fig_5} and \ref{fig_6} and present 
the deduced optical properties of the dust grains for the PA sector 
45$^{\circ}$ - 90$^{\circ}$, which includes the brightest part of the 
luminous arc, and PA sector 315$^{\circ}$ - 360$^{\circ}$, which covers 
the diffuse echo ring observed with the highest S/N. They are shown in 
Figure \ref{fig_9} for F475W and Figure \ref{fig_10} for F606W, both on 
+277 d. In these diagrams, rectangular coordinates $x$ and $y$ are 
replaced with polar coordinates $\rho$ and $\phi$, and the abscissa 
corresponds to the physical distances in the plane of the sky. The left 
ordinate represents the quantity $\omega C_{ext} n_d (\rho, \phi, t)$, 
which is determined by the optical properties of the dust grains.  The 
right ordinate shows $\omega C_{ext} n_d dz = \omega \tau$, where $\tau$ 
is the optical depth of the dust mapped onto a single pixel. 
By looking at the entire echo profile, we found that a major part of the 
luminous-arc echo spreads over 45$^{\circ}$ - 180$^{\circ}$ in PA, and the 
diffuse echo ring attained the highest S/N over 270$^{\circ}$-360$^{\circ}$ 
in PA.  For completeness, on-line Figures \ref{sfig_1} and
\ref{sfig_2} present the same diagrams over the entire eight bins in PA. 

We applied a Galactic extinction model with $R_V$=3.1 to the 
scattering materials and compare the reproduced extinction properties 
with the extinction along the DLOS. Discrepancies between the derived 
quantities and the assumed model will indicate that the extinction 
properties of the scattering dust are different from the Milky Way dust 
with $R_V$=3.1. 
For each photometric bandpass its pivot wavelength was used in 
interpreting the parameters from dust models. 
The extinction curve is obtained from \citet{Weingartner_Draine_2001} 
and \citet{Draine_2003a, Draine_2003b}
\footnote[1]{ftp://ftp.astro.princeton.edu/draine/dust/
mix/kext\_albedo\_WD\_MW\_3.1\_60\_D03.all}. 
For $C_{ext}$, 
the extinction cross section per hydrogen nucleon H, we adopted
$5.8\times10^{-22}$ cm$^2$/H for F475W, and $4.4\times10^{-22}$
cm$^2$/H for F606W; for the scattering 
phase function, we adopted $g=0.555$ for F475W, and $g=0.522$ for F606W, 
and $n_d$ is the H volume number density in units of cm$^{-3}$. 

For a uniform dust distribution in the x/y direction (in the plane 
of the sky), integrating $\omega \tau$ over each position angle will
provide a rough estimate of the product of the total optical depth
and the scattering albedo, which is the main value added by the
separate analysis of light echoes.  We applied the same extinction
measured along the DLOS to the scattered light echoes and calculated 
the optical depth of the materials from scattering.  This is 
labeled by the red text in the upper right of each panel of Figures 
\ref{fig_9} and \ref{fig_10}.  The inhomogeneity of the ISM in M82 
has small scales as is indicated by the rapid variability of the 
strength of the echo with PA along the rings as well as with time.  
The optical depth along the DLOS has been measured by  
\citet{Foley_etal_2014} around maximum light as 
$\tau_B$ = 3.11(0.18) and $\tau_V$ = 1.91(0.17) based on $A_B$ = 3.38(20),  
$A_V$ = 2.07(18) and using the relation $A_{\lambda} = -2.5 log_{10} 
(e)\tau_{\lambda} ^{ext} = 1.086 \tau_{\lambda}^{ext}$.

The hydrogen column number density along the line of sight is 
$n_H = \int_{LOS} n_d(z) dz$.  Therefore, $n_H$ can be obtained by 
dividing the total optical depth per bin in position angle by $\omega
C_{ext}$ (Figure \ref{fig_9} for F475W and Figure \ref{fig_10} for 
F606W). For example, for F475W and +277 d, the maximum value of 
$\omega \tau (\rho,\phi,t)$ in the luminous arc was observed to be 
around 0.58. Using $\omega \sim 0.65$ for the Milky-Way dust model 
with $R_V$ = 3.1 given by \citet{Weingartner_Draine_2001}, {$n_H$ can 
be estimated to be} $n_H = 0.58 / \omega C_{ext} = 0.58 / (0.65 \times 
5.8 \times 10^{-22}$ cm$^2$/H) $\sim1.5\times 10^{21}$ H cm$^{-2}$ 
in the bin which shows the densest part of the dust slab producing 
the luminous arc echo. 
This is $\sim$15 times denser than the scattering material in the
foreground of the Type-II plateau SN~2008bk \citep{VanDyk_2013}, for
which the visual extinction of the dust responsible for the echo is
$A_V \approx$ 0.05.  It is also $\sim4$ times denser than the ISM in
the foreground of the Type II-plateau SN~2012aw \citep{VanDyk_2015},
for which the dust extinction in the SN environment responsible for
the echo is consistent with the value that was estimated from
observations of the SN itself at early times, i.e., $A_V$=0.24. 

Figure \ref{fig_12} presents the three-dimensional dust distribution
estimated for SN~2014J. {Data-points show the number densities
as derived} from two iso-delay paraboloids. Scattering materials 
producing the luminous arc and the diffuse echo, respectively were 
mapped out at epochs +277 d (inner layer) and +416 d (outer layer). 

\subsection{Scattering wavelength dependence of the ISM}
From the scattering properties of the dust, its optical properties can 
be estimated by comparing the quantity $\omega C_{ext} n_d$ derived 
for F475W and F606W.  Figure \ref{fig_11} presents the division of the 
profiles of Figure \ref{fig_9} by Figure \ref{fig_10}. This yields the 
wavelength dependence of the extinction cross-section.  As
the ordinate of Figure \ref{fig_11} we use $\omega \tau_{F475W} /
\omega \tau_{F606W}$.  Overplotted histograms show (in red) the number
density of the scattering material derived from the strength of the echoes 
in F475W. The horizontal gray dashed lines mark the value of
$\tau_{F475W} / \tau_{F606W} = A_{F475W} / A_{F606W} = $1.66, 1.30, and 
1.19 for Milky Way-like dust with $R_V = $1.4, 3.1, and 5.5, respectively,  
according to the algorithm determined by \citet{Cardelli_etal_1989}. 
Similar diagrams over the entire eight bins of PA are shown by Figure 
\ref{sfig_3} in the electronic version. 

Plausible estimates of $\omega \tau_{F475W} / \omega \tau_{F606W}$ can 
only be made in high-S/N regions of the echoes. In the left panel of 
Figure \ref{fig_11}, the luminous arc at $\rho$ = 10$\sim$11 pc has an 
average value $\omega \tau_{F475W} / \omega \tau_{F606W} \sim$1.7 
(dimensionless), shown by the black histograms. For the diffuse structure,
the right panel indicates an average value $\sim$1.3. This difference 
in the wavelength dependence measured from the scattering optical depth 
indicates that the size of the grains in the thin dust 
slab producing the luminous arc are different from the grains in the 
foreground extended dust cloud producing the diffuse echo.  While this 
difference is significant, one should be cautious about the inferred 
absolute values of $R_V$ in this approach, considering the low 
signal-to-noise ratio and the large uncertainties. 

\section{Discussion}
The diffuse echo component favors a higher $R_V$ than the luminous arc, 
corresponding to a less steep wavelength dependence of the extinction 
in the diffuse echo compared to the luminous arc. In general terms, 
this implies that the grains in the dust slab producing the luminous 
arc are smaller than those in the extended, diffuse ISM. The $R_V$ value 
measured from the diffuse echo at $\rho \sim 10 - 14$ pc to the position 
of SN~2014J, i.e., $R_V \sim 3$, is close to that found by 
\citet{Hutton_etal_2015} by modelling the attenuation law based on
near-ultraviolet and optical photometry of M82 at large.
Accordingly, the dust grains 
in the extended foreground ISM producing the diffuse echo ring are 
similar in size to those in the Milky Way. 
Extinction in the luminous 
arc, however,  favors a smaller $R_V$ value, similar to the extinction 
law deduced from the SN itself, represented by $R_V \sim 1.4$. This 
similarity indicates that the grain size distribution in the slab of 
ISM producing the luminous arc is similar to the ISM responsible for 
the extinction measured towards the SN at early epochs. 

The optical depth due to light scattered by the ISM can be estimated 
as follows. If they have similar properties as 
Milky Way-like dust with $R_V$ = 3.1, $\tau_{F475W}$ ranges from 
0.3 at PA 225$^\circ$ - 270$^\circ$, covering part of the diffuse ring, 
to 0.9 at PA 45$^\circ$ - 90$^\circ$, where the luminous arc is brightest.
These optical depths are smaller than that along the DLOS. 
One possible explanation for the discrepancy can be an 
overestimate of the degree of forward scattering. At +277 d, the 
scattering angle is $\sim2.6^{\circ}$ for the luminous arc-producing dust. 
A dramatic increase in forward scattering occurs with increasing grain
size while smaller grains scatter light more isotropically, leading to 
a smaller value of the phase function, {see Chapter 5 of 
 \citet{Van_de_Hulst_1957}}. Therefore,
to produce a light echo of the same strength, 
smaller dust grains in the ISM responsible for the luminous arc  
require a higher optical depth than larger Milky Way-like dust grains do. 

To illustrate the dependence of the degree of forward scattering on 
the optical depth, we investigate the Heyney-Greenstein phase function 
characterizing the angular distribution of scattered light intensity as 
shown by Equation \ref{eqn_phasefunction}. Figure \ref{fig_13} demonstrates 
the fraction of scattered light at small scattering angle, i.e., 
$2.6^{\circ}$ as a function of scattering asymmetry factor, $g$. In this 
figure, values of $g=0.439$ and $g=0.345$ are indicated for astronomical 
silicate and graphite grains with radius of $0.1$ micron according to 
calculations based on \citet{Draine_Lee_1984} and \citet{Laor_Draine_1993}. 

When the grains are significantly smaller than the wavelength of light, 
the classical Rayleigh scattering limit is reached.  The asymmetry factor
for Rayleigh scattering is $g=0$, and the phase function becomes unity,
indicating no directional preference of scattering.  This is the case for
the luminous arc while the phase function has a value of 7.8 for
Milky-Way dust with $R_V$ = 3.1. This means that the optical 
depth calculated for the case of Rayleigh scattering is 7.8 times larger 
than for Milky-Way dust with $R_V$ = 3.1. The densest part of the
scattering material will attain a value of $\sim$ 7.0 
in F475W, significantly larger than the optical depth measured along the 
DLOS. On the other hand, for larger grains the asymmetry 
factor $g$ approaches unity, and the efficiency of forward scattering
increases substantially.  

The grain size distribution in the extinction-producing material
towards SN~2014J itself is similar to that of the luminous 
arc-producing material, as inferred from the similarity of $R_V$ 
found in both of the two ISM components. Considering this low 
$R_V$ and the lower optical depth found in the 
scattering material responsible for the luminous arc, we infer that 
these scattering materials are also responsible for the extinction 
towards SN~2014J. 
{Our result is consistent with the relationship between the host
galaxy extinction $A_V$ and their measured $R_V$ \citep{Mandel_etal_2011}, 
which for SNe with low extinction, 
$A_V \lesssim 0.4$, $R_V \approx 2.5-2.9$ is favored, while at high 
extinction, $A_V \gtrsim 1$, low values of $R_V \textless2$ are favored.} 
Due to the lack of knowledge about the detailed 
distribution and optical properties of the dust in M82, we cannot 
rule out the possibility that the different extinctions along the 
scattering line of sight of the materials and the DLOS may 
partly also be caused by a denser ISM along the DLOS. The extinction 
along the DLOS may also be due to dust at small foreground distances 
which would produce light echoes too close to the SN to be detected. 
Additionally, it is possible that the extinction can 
be generated by interstellar dust clouds placed too far in front of 
the SN. Recall Equations \ref{eqn_r} and \ref{eqn_lumi}, 
the luminosity of the light echo resulting from a dust 
slab intersecting the DLOS decreases as $1/r$ (where $r$ is the 
distance between the SN and the dust slab). {Considering numerous Na, 
Ca, and K features have been seen along the DLOS \citep{Patat_etal_2015}}, 
we cannot rule out 
the possibility that there are dust clouds placed more than 500 pc 
away from the SN and can hardly be detected in current images. 

The presented light-echo model is necessarily only a simplified 
approximation of reality. Our model attempts to reproduce the optical 
depth of the scattering material over a projected area of $\sim$ 40 pc 
$\times$ 40 pc in the plane of the sky, {and compares it to the} 
optical depth measured for the DLOS.  One major source of uncertainty 
is the assumption of single scattering 
\citep{Wood_etal_1996, Patat_2005}.  In view of the large extinction 
measured towards SN~2014J, a Monte Carlo simulation with various grain 
size distributions should give a better representation of the 
real scattering process. Another uncertainty results from the usage of the 
extinction measured along the DLOS around maximum light also for the 
echo-producing material.  Additionally, the assumption of Galactic 
$R_V$ values may not be realistic for M82.

\section{Summary}
The geometric and photometric evolution of resolved light echoes
around SN~2014J was monitored with {\it HST}.  Two main constituents
were found.  From a luminous arc, a discrete slab of dust 
was inferred at a foreground distance of 228$\pm$7pc.  In addition, a
resolved, diffuse ring-like light echo implies that another foreground
ISM component is widely distributed over distances of $\sim$ 100-500 pc.  
If the scattering material suffers the same extinction as along the
DLOS, the densest part has a number density of $\sim$ 1.5 $\times
10^{21}$ cm$^{-2}$, based on a single-scattering-plus-attenuation
approach.  The scattering material is unevenly distributed with PA. The 
wavelength dependence of the scattering optical depth is steeper in 
the luminous arc than in the diffuse ring.  The former favors a 
small $R_V \sim$ 1.4 as also measured along the DLOS, and the latter is
more consistent with a 'normal' $R_V \sim$ 3. This suggests that the 
average grain size is smaller in the ISM responsible 
for the luminous arc, and the more widely distributed scattering 
materials have average properties similar to Milky Way-like dust. 
{This study reveals the $R_V$ fluctuation of the extragalactic dust 
on parsec scales. We deduce that systematically steeper extinction laws 
towards Type Ia SNe do not have to represent the average behavior of 
the extinction law in the host galaxy.} 

The optical depth of the scattering material estimated from the scattering
properties of Milky-Way-like dust with $R_V$ = 3.1 is smaller than the
optical depth measured along the DLOS.  The optical depth along the
DLOS is better reproduced with smaller grains as also indicated for
the dust slab responsible for the luminous arc. This suggests that an
extension of this dust slab, or a separate cloud with similar
properties, is also responsible for the extinction towards
SN~2014J. More data will be collected {in future observing campaigns 
that will help} additionally characterize the extinction laws measured
within different light echo components. 
Further constraints on the amount and properties of the circumstellar 
and interstellar material from polarimetry and very late-time photometry 
will be discussed in future work. 

\acknowledgments We are grateful to Peter Lundqvist and Anders Nyholm
for providing the late-time spectrum of SN~2014J.  We also thank Jian
Gao, Bi-wei Jiang, Kevin Krisciunas, Armin Rest, and Nicholas Suntzeff
for helpful discussions.  The supernova research by Y. Yang,
P. J. Brown, and L. Wang is supported by NSF grant AST-0708873.
P. J. Brown was partially supported by a Mitchell Postdoctoral
Fellowship.  Y. Yang and M. Cracraft also acknowledge support from
NASA/STScI through grant HST-GO-13717.001-A.  L. Wang is supported by
the Strategic Priority Research Program "The Emergence of Cosmological
Structures" of the Chinese Academy of Sciences, Grant No. XDB09000000.
L. Wang and X. Wang are supported by the Major State Basic Research
Development Program (2013CB834903), and X. Wang is also supported by
the National Natural Science Foundation of China (NSFC grants 11178003
and 11325313).

\begin{figure}[!htbp]
\epsscale{1}
\plotone{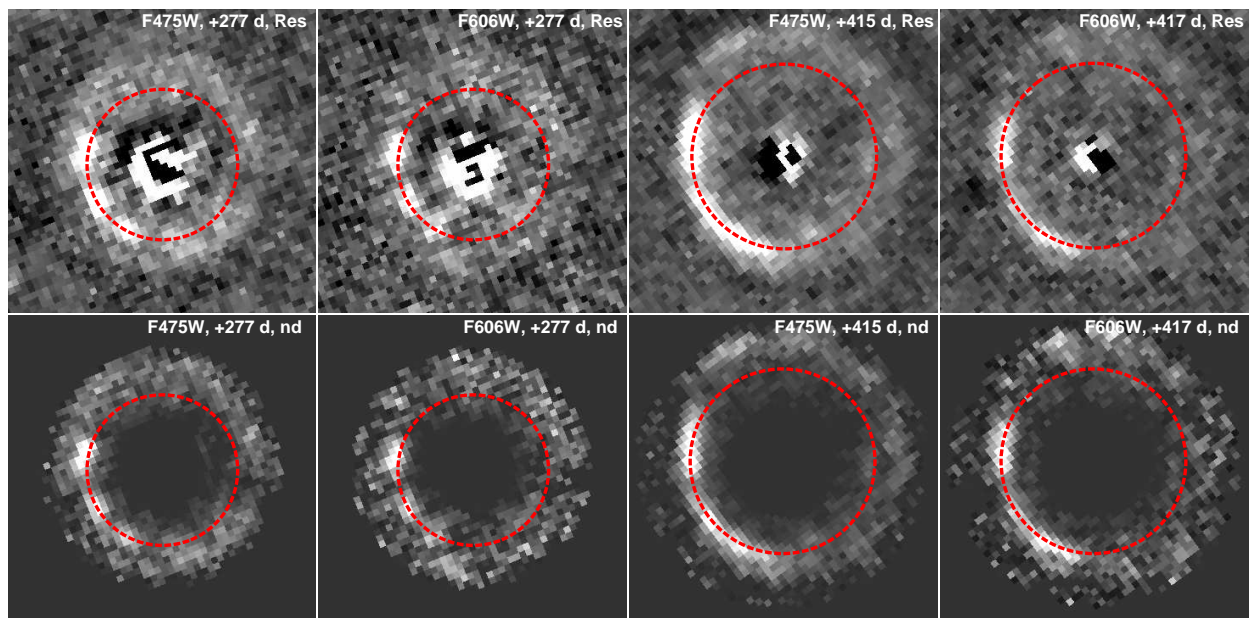}
\caption{Intensity maps of the backgound- and PSF-subtracted images
  (labeled `Res') and scaled volume number-density maps (`nd') showing 
  the relative column density calculated from the flux and location 
  in space of each pixel.  North is up and east is left.  Epoch and
  passband of the observations are labeled.  Dashed circles trace the
  dust slab at $z\sim$228 pc, which is responsible for the luminous
  echo arc appearing with different diameter at different epochs.
  Overdensities can be identified at PAs
  60$^{\circ}$ -- 180$^{\circ}$ along the lunimous arc and also at
  PAs 0$^{\circ}$ -- 60$^{\circ}$ and larger radii in F475W and
  F606W +416/+417 d.
\label{fig_8}} 
\end{figure}

\begin{figure}[!htbp]
\epsscale{0.8}
\plotone{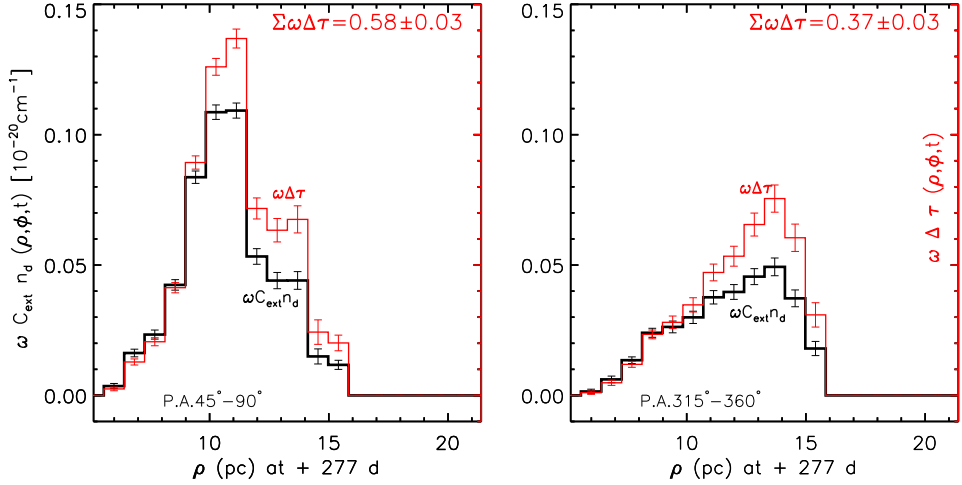}
\caption{
  Radial profiles at different PAs (as labeled) of optical
  properties of the scattering material. The calculations are based on 
  the density map (transformed from the residual image) in passband 
  F475W on +277 d. The left panel shows the luminous arc echo, 
  and the right panel presents the diffuse ring echo. The $x$-axis 
  shows the physical distances in the plane of the sky ($\rho$-direction). 
  Black histograms represent $\omega C_{ext} n_d (\rho, \phi, t)$ in 
  units of $10^{-20}cm^{-1}$ as shown on the left ordinate and can 
  be used to infer the volume densities. Red histograms represent 
  the dimensonless $\omega C_{ext} n_d dz = \omega \tau$ and share 
  the tick marks of the left ordinate, from which 
  column number densities can be deduced. The optical depth of the 
  dust mapped onto a single pixel gives $\tau$. 
\label{fig_9}} 
\end{figure}

\begin{figure}[!htbp]
\epsscale{0.8}
\plotone{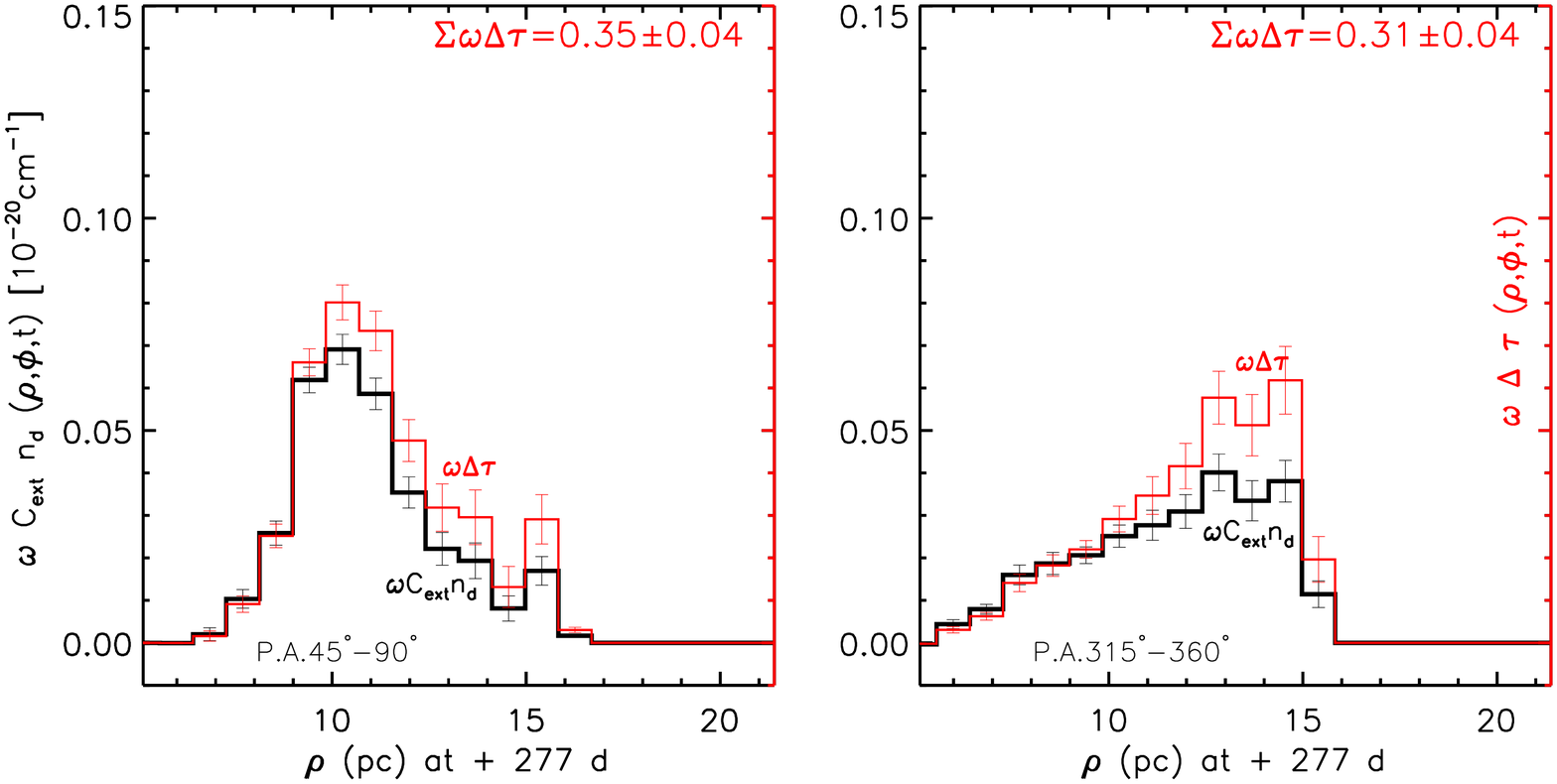}
\caption{Same as Figure \ref{fig_9} except for F606W. 
\label{fig_10}} 
\end{figure}

\begin{figure}[!htbp]
\epsscale{0.8}
\plotone{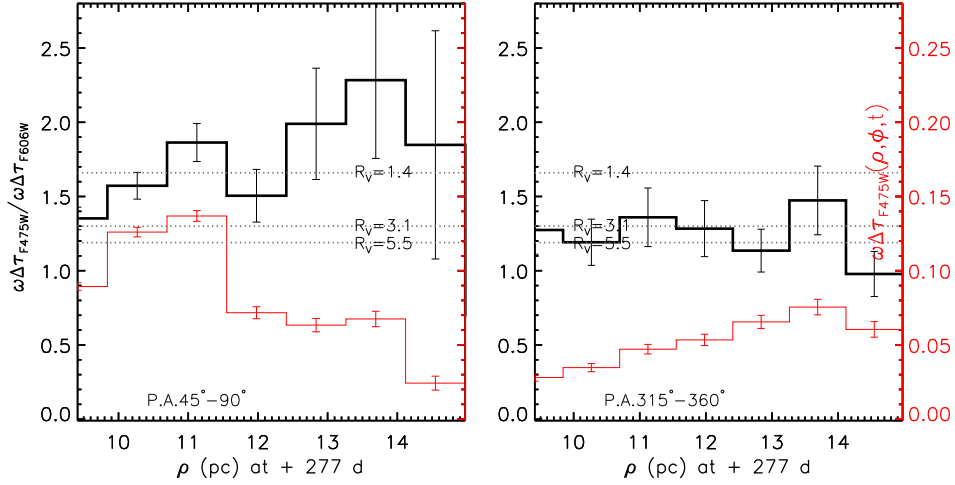}
\caption{
  Radial run of the wavelength dependence of the scattering material
  characterized by $\omega \tau_{F475W}  / \omega \tau_{F606W}$ on 
  +277 d, shown by the black histograms. Red histograms represent 
  the dimensionless quantity $\omega C_{ext} n_d dz = \omega \tau$,
  which is a measure of the 
  strength of the echoes. The abscissa measures the physical 
  distances (in pc) in 
  the plane of the sky. The upper, middle, and lower horizontal 
  dashed lines represent the values calculated for Milky Way
  extinction laws with $R_V$ 
  = 1.4, 3.1, and 5.5, respectively. The left panel includes the 
  luminous arc echo at $\rho$ = 10$\sim$11 pc and 
  $\omega \tau_{F475W} / \omega \tau_{F606W}$ $\sim$1.7. The right 
  panel presents the diffuse ring echo, exposing a different 
  wavelength dependence of scattering with $\omega \tau_{F475W} 
  / \omega \tau_{F606W} \sim 1.3$.
\label{fig_11}} 
\end{figure}

\begin{figure}[!htbp]
\epsscale{1}
\plotone{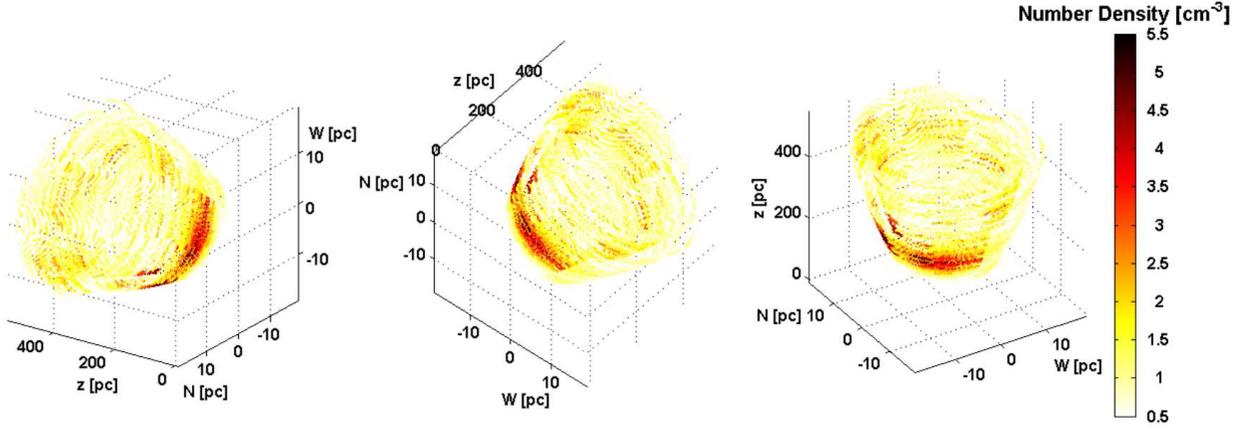}
\caption{The three-dimensional dust distribution derived from the
  light echoes around SN~2014J. From left to right, the vertical axis
  corresponds to the directions East-West, North-South, and the DLOS
  (z).  The color encoding of the number density of the dust is 
  indicated by the vertical bar. The measurements map out density 
  along iso-delay parabloids as schematically depicted in 
  Figure~\ref{fig_1}. They correspond to epochs +277 d and +416 d 
  and {are too close to one another to appear separately}. 
\label{fig_12}} 
\end{figure}

\begin{figure}[!htbp]
\epsscale{0.6}
\plotone{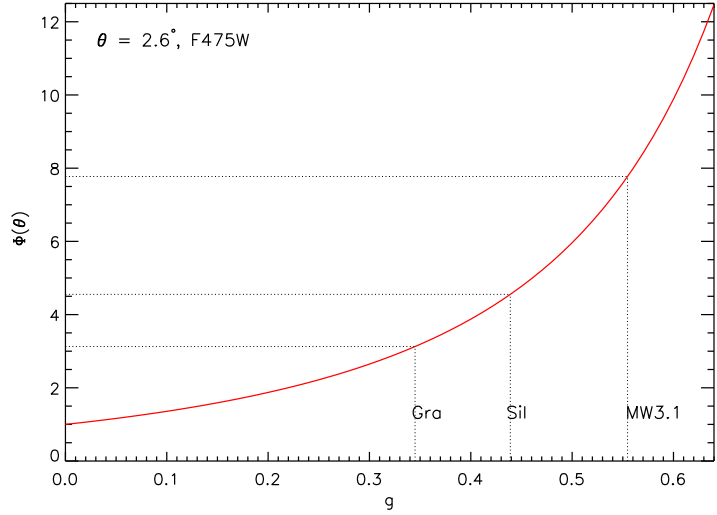}
\caption{
  Fraction of light scattered at the small 
  angle of $2.6^{\circ}$ as a function of the value of the phase scattering
  function, $g$, as calculated from 
  Equation \ref{eqn_phasefunction}.  MW3.1, Gra, and Sil represent the 
  $g$ factors for Milky-Way dust with $R_V = 3.1$, graphite spheres with 
  radius 0.1 $\mu m$, and "astronomical silicate" spheres with 
  radius 0.1 $\mu m$. 
\label{fig_13}} 
\end{figure}

\begin{deluxetable}{ccccc}
\tablewidth{0pc}
\tabletypesize{\scriptsize}
\tablecaption{Spatially resolved supernova light echoes}
\tablehead{
\colhead{SN} \vspace{-0.0cm}  &  \colhead{Type}  &  \colhead{Host}  & \colhead{Distance} & \colhead{References}  \\ \vspace{-0.0cm} 
 & & \colhead{Galaxy} & \colhead{(Mpc)} & \colhead{$^a$}}
\startdata
1987A  &  II-Peculiar  &  LMC       &  0.05  & 1, 3, 16, 17, 23 \\
1991T  &  Ia 91T-like  &  NGC 4527  &  15.2  & 11, 12 \\
1993J  &  IIb          &  M81       &  3.6   & 6, 13  \\
1995E  &  Ia           &  NGC 2441  &  49.6  & 10 \\
1998bu &  Ia           &  M96       &  9.9   & 2  \\
1999ev &  II-P         &  NGC 4274  &  9.9   & 7  \\
2002hh &  II-P         &  NGC 6946  &  5.5   & 8, 22  \\
2003gd &  II-P         &  M74       &  9.5   & 14, 18 \\
2004et &  II-P         &  NGC 6946  &  5.5   & 9  \\
2006X  &  Ia           &  M100      &  15.9  & 21 \\
2007af &  Ia           &  NGC 5584  &  22.5  & 5  \\
2008bk &  II-P         &  NGC 7793  &  3.7   & 19 \\
2012aw &  II-P         &  M95       &  10.0  & 20 \\
2014J  &  Ia           &  M82       &  3.5   & 4  \\
2016adj&  IIb          &  NGC 5128  &  3.7   & 15 \\
\enddata
\tablenotetext{a}{
(1) \citet{Bond_etal_1990}, 
(2) \citet{Cappellaro_etal_2001}
(3) \citet{Crotts_1988}
(4) \citet{Crotts_2015}
(5) \citet{Drozdov_etal_2015}
(6) \citet{Liu_etal_2003}
(7) \citet{Maund_etal_2005}
(8) \citet{Meikle_etal_2006}
(9) \citet{Otsuka_etal_2012}
(10) \citet{Quinn_etal_2006}
(11) \citet{Schmidt_etal_1994}
(12) \citet{Sparks_etal_1999}
(13) \citet{Sugerman_etal_2002}
(14) \citet{Sugerman_2005}
(15) \citet{Sugerman_etal_2016}
(16) \citet{Suntzeff_etal_1988}
(17) \citet{Spyromilio_etal_1995}
(18) \citet{VanDyk_etal_2006}
(19) \citet{VanDyk_2013}
(20) \citet{VanDyk_2015}
(21) \citet{Wang_etal_2008}
(22) \citet{Welch_etal_2007}
(23) \citet{Xu_etal_1994}}
\label{Table_1}
\end{deluxetable}

\begin{deluxetable}{ccccccccc}
\tablewidth{0pc}
\tabletypesize{\scriptsize}
\tablecaption{Log of observations of SN~2014J with {\it HST} WFC3/UVIS and ACS/WFC POLV \label{Table_2}}
\tablehead{
\colhead{HST} \vspace{-0.3cm}  &  &  & \colhead{Date of $1^{st}$ Obs.} &  \colhead{Exp. Time}  & \colhead{Epoch$^a$} & \colhead{Date of $2^{nd}$ Obs.} & \colhead{Exp. Time} & \colhead{Epoch$^a$}  \\ \vspace{-0.3cm} 
 & \colhead{Filter} & \colhead{polarizer} & & & & & & \\ 
\colhead{Camera} &  &  & \colhead{(UT - 2014)} &  \colhead{(s)}  & \colhead{(Days)} & \colhead{(UT - 2015)} & \colhead{(s)} & \colhead{(Days)}   }
\startdata
WFC3/UVIS$^{b}$ &  F438W & N/A      & 09-05 19:12:57 & 8$\times$64  & 215.8         & 02-02 05:24:41 & 12$\times$128 & 365.2 \\
                &  F555W & N/A      & 09-05 19:29:44 & 4$\times$64  & 215.8         & 02-02 05:06:06 & 12$\times$32  & 365.2 \\
                &  F555W & N/A      & 09-05 22:05:11 & 8$\times$32  & 215.9         &          N/A        &      N/A      &  N/A  \\
                &  F814W & N/A      & 09-05 20:32:05 & 8$\times$64  & 215.9         &          N/A        &      N/A      &  N/A  \\
\hline
ACS/WFC$^{c}$   &  F475W & POL0V   & 11-06 00:24:42 & 3$\times$130  & 276.5         & 03-25 01:56:17 & 3$\times$400  & 415.6 \\
                &  F475W & POL120V & 11-06 00:42:24 & 3$\times$130  & 276.5         & 03-25 03:22:43 & 3$\times$400  & 415.6 \\
                &  F475W & POL60V  & 11-06 01:00:03 & 3$\times$130  & 276.5         & 03-25 03:53:40 & 3$\times$400  & 415.7 \\
                &  F606W & POL0V   & 11-06 01:18:11 & 2$\times$40   & 276.6         & 03-27 10:17:38 & 3$\times$60   & 417.9 \\
                &  F606W & POL120V & 11-06 01:59:48 & 2$\times$40   & 276.6         & 03-27 11:10:48 & 3$\times$60   & 418.0 \\
                &  F606W & POL60V  & 11-06 02:13:58 & 2$\times$40   & 276.6         & 03-27 11:30:17 & 3$\times$60   & 418.0 \\
                &  F775W & POL0V   & 11-06 02:23:28 & 2$\times$30   & 276.6         & 03-27 11:50:26 & 3$\times$20   & 418.0 \\
                &  F775W & POL120V & 11-06 02:37:21 & 1$\times$55   & 276.6         & 03-27 12:58:00 & 3$\times$20   & 418.0 \\
                &  F775W & POL60V  & 11-06 02:41:46 & 1$\times$55   & 276.6         & 03-27 13:02:17 & 3$\times$20   & 418.0 \\
\enddata
\tablenotetext{a}{Days after B maximum on 2014 Feb. 2.0 (JD 2 456 690.5).}
\tablenotetext{b}{Observations result from {\it HST} WFC3/UVIS, program (\#13626; PI: Crotts)}
\tablenotetext{c}{Observations result from {\it HST} ACS/WFC, program (\#13717; PI: Wang)}
\end{deluxetable}

\begin{deluxetable}{cccccc}
\tablewidth{0pc}
\tabletypesize{\scriptsize}
\tablecaption{HST photometry of SN~2014J and light echoes (total echo profile) \label{Table_3}}
\tablehead{
\colhead{t$^a$} \vspace{-0.3cm}& \colhead{F438W$_{SN}$} & \colhead{F555W$_{SN}$} & \colhead{F814W$_{SN}$} & \colhead{F438W$_{LE}$} & \colhead{F555W$_{LE}$} \\ \vspace{-0.1cm} 
}
\startdata
215.8      &  17.610$\pm$0.016  &  16.446$\pm$0.011  &  15.301$\pm$0.011  &  22.05$\pm$0.36 &  21.12$\pm$0.06 \\
365.3      &  19.735$\pm$0.011  &  18.715$\pm$0.013  &  N/A$^b$           &  21.53$\pm$0.13 &  20.87$\pm$0.06 \\
\hline
t$^a$      &  F475W$_{SN}$      &  F606W$_{SN}$      &  F775W$_{SN}$      &  F475W$_{LE}$   &  F606W$_{LE}$   \\
276.5      &  17.467$\pm$0.002  &  17.343$\pm$0.002  &  16.354$\pm$0.005  &  21.16$\pm$0.03 &  20.73$\pm$0.08 \\
415.6$^c$  &  19.568$\pm$0.002  &  19.516$\pm$0.004  &  17.888$\pm$0.008  &  21.37$\pm$0.02 &  20.98$\pm$0.05 \\

\enddata
\tablenotetext{a}{Days after $B$ maximum, 2014 Feb. 2.0 (JD 245 6690.5). }
\tablenotetext{b}{SN~2014J was not observed in F814W at +365 d. }
\tablenotetext{c}{+417.9 d for F606W, +418.0 d for F775W. }
\end{deluxetable}

\begin{deluxetable}{lcccccc}
\tablewidth{0pc}
\tabletypesize{\scriptsize}
\tablecaption{Geometric properties of (unresolved) luminous-arc light-echo (LE) components \label{Table_4}}
\tablehead{
\colhead{LE} \vspace{-0.3cm}  &  \colhead{Epoch$^1$}  &  \colhead{Angular Radius}       &  \colhead{Offset}      &  \colhead{Foreground Distance}  &  \colhead{Projected Radius} &  \colhead{Scattering Angle} \\ \vspace{-0.3cm} 
& & & & & & \\ 
\colhead{\#}                  &  \colhead{(Day)}      &  \colhead{$\alpha$ ($\arcsec$)} &  \colhead{($\arcsec$)} & \colhead{$z$ (pc)}       &  \colhead{$\rho$ (pc)} &  \colhead{$\theta$ ($^{\circ}$)} }
\startdata
Arc                           & 215.8                 &  0.539$\pm$0.020   &   0.009$\pm$0.014    & 234.6$\pm$18.2  &  9.22$\pm$0.36  &  2.25$\pm$0.20  \\
                              & 276.5                 &  0.599$\pm$0.014   &   0.006$\pm$0.015    & 226.3$\pm$11.8  & 10.25$\pm$0.27  &  2.60$\pm$0.15  \\
                              & 365.3                 &  0.689$\pm$0.020   &   0.011$\pm$0.014    & 226.4$\pm$14.1  & 11.79$\pm$0.37  &  2.98$\pm$0.21  \\
                              & 415.6                 &  0.735$\pm$0.012   &   0.012$\pm$0.010    & 226.6$\pm$9.0   & 12.58$\pm$0.25  &  3.18$\pm$0.14  \\
\enddata
\tablenotetext{1}{Days after $B$ maximum on 2014 Feb. 2.0 (JD 245 6690.5). }
\end{deluxetable}

\begin{figure}[!htbp]
\epsscale{0.8}
\plotone{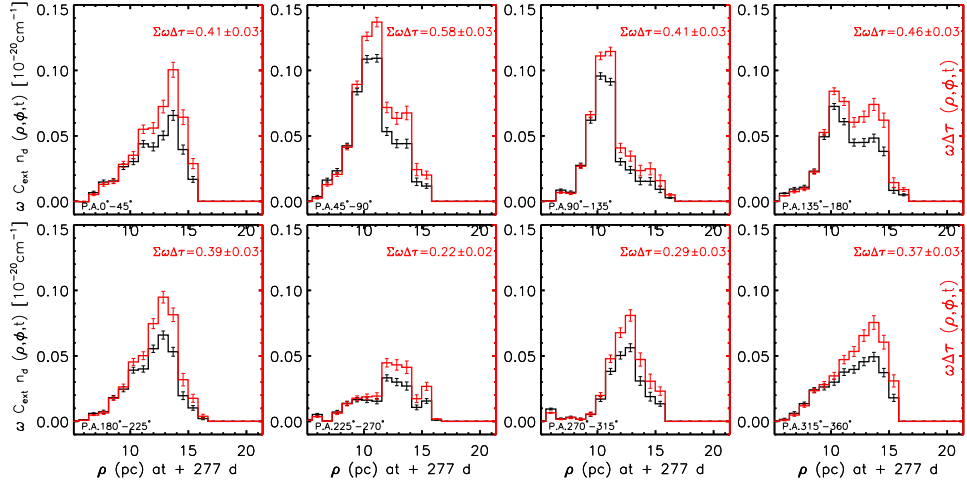}
\caption{Radial profiles at different PAs (as labeled) of optical
  properties of the scattering material. The calculations are based on 
  the density map (transformed from the residual image) in passband 
  F475W on +277 d. Black histograms represent $\omega C_{ext} n_d
  (\rho, \phi, t)$ in units of $10^{-20}cm^{-1}$ as shown on the left 
  ordinate and can be used to infer the volume densities. Red 
  histograms represent the unitless $\omega C_{ext} n_d dz = \omega 
  \tau$ and share the same tick marks as the left ordinate, which 
  can be used to infer the column number densities. $\tau$ is the 
  optical depth of the dust mapped onto a single pixel. 
\label{sfig_1}} 
\end{figure}

\begin{figure}[!htbp]
\epsscale{0.8}
\plotone{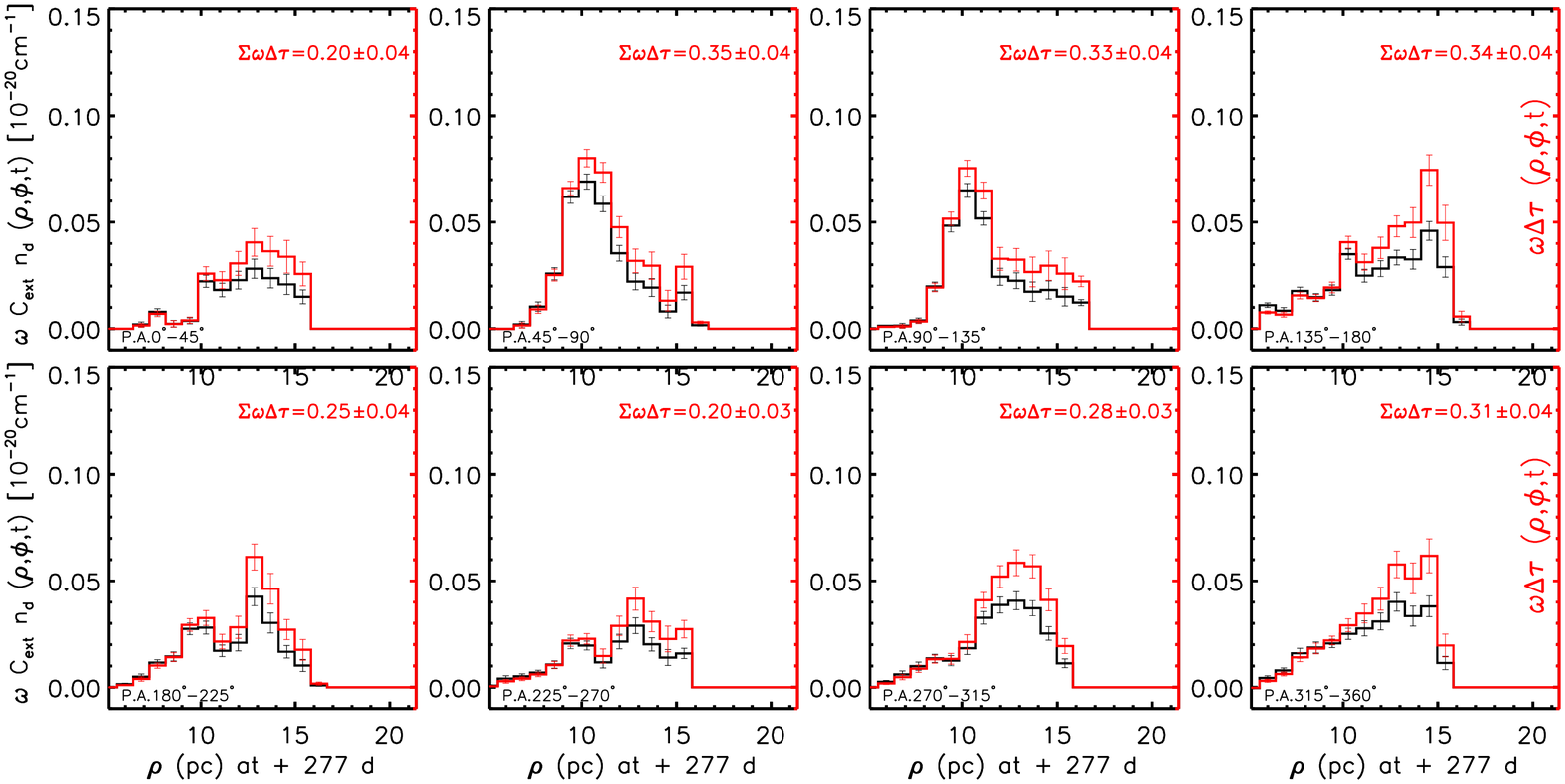}
\caption{Same as Figure \ref{sfig_1} except for F606W. 
\label{sfig_2}} 
\end{figure}

\begin{figure}[!htbp]
\epsscale{0.8}
\plotone{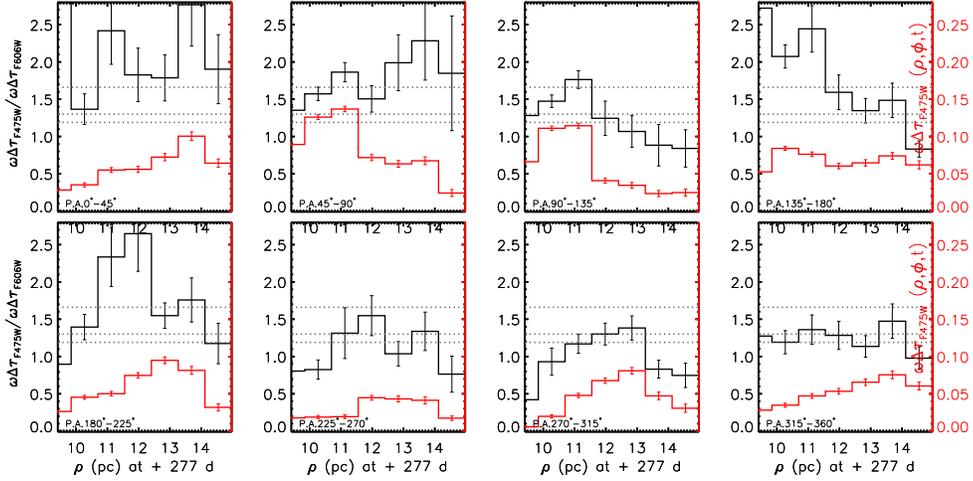}
\caption{
  Radial run of the wavelength dependence of the scattering material 
  characterized by $\tau_{F606W}$ on +277 d.  The abscissa measures 
  the physical distances (in pc) in the plane of the sky.  Each panel 
  shows a different bin in position angle of width $45^{\circ}$ (as 
  labeled). The upper, middle, and lower horizontal dashed lines 
  represent the values calculated for Milky-Way dust with $R_V$ = 1.4, 3.1, and 
  5.5, respectively. The luminous arc in PA bins from 45$^{\circ}$ to 
  180$^{\circ}$ appear at $\rho$ = 10$\sim$11 pc and $\omega \tau_{F475W} / 
  \omega \tau_{F606W}$ $\sim$1.7. Diffuse structures at large PAs 
  expose a different wavelength dependence on scattering because 
  $\omega \tau_{F475W} / \omega \tau_{F606W}\sim 1.3$. 
\label{sfig_3}} 
\end{figure}

\end{document}